\documentclass[10pt,twocolumn,twoside]{IEEEtran}

\usepackage{cite}
\usepackage{amsmath}
\usepackage{amssymb}
\usepackage{fontenc}
\usepackage{color}

\newtheorem{theorem}{Theorem}

\usepackage{algorithm}
\usepackage{algorithmic}
\usepackage{subfigure}
\usepackage{graphicx}%[pdftex]
\usepackage[caption=false,font=footnotesize]{subfig}
\usepackage{url}
\usepackage{bm}
\usepackage{booktabs}
\usepackage{balance}
\usepackage{xcolor}
\usepackage{multirow}
\usepackage{booktabs}
\usepackage{balance}
\usepackage{epstopdf}
\usepackage{threeparttable}

\newcommand{\tabincell}[2]{\begin{tabular}{@{}#1@{}}#2\end{tabular}}

\begin{document}

\title{Unsupervised Recurrent Federated Learning for Edge Popularity Prediction in Privacy-Preserving Mobile Edge Computing Networks}

\author{Chong~Zheng,~\IEEEmembership{Graduate~Student~Member,~IEEE},
        Shengheng~Liu,~\IEEEmembership{Senior~Member,~IEEE},
        Yongming~Huang,~\IEEEmembership{Senior~Member,~IEEE},
        Wei~Zhang,~\IEEEmembership{Fellow,~IEEE},
        and~Luxi~Yang,~\IEEEmembership{Senior~Member,~IEEE}

\thanks{Manuscript received XXX XX, XXXX; revised XXX XX, XXXX; accepted XXX XX, XXXX. This work was supported in part by the National Natural Science Foundation of China under Grant Nos. 62001103 and 61720106003, and the National Key R\&D Program of China under Grant No. 2018YFB1800801. This work was also supported in part by the Australian Research Council's Project funding scheme under LP160101244. Part of this work has been accepted for presentation at the IEEE Global Communications Conference (GLOBECOM): Machine Learning for Communications Symposium, Madrid, Spain, December 2021 \cite{ZhengCGlo21}. (Corresponding authors: Y.~Huang; S.~Liu.)}

\thanks{C.~Zheng, S.~Liu, Y.~Huang, and L.~Yang are with the National Mobile Communications Research Laboratory, School of Information Science and Engineering, Southeast University, Nanjing 210096, China, and also with the Purple Mountain Laboratories, Nanjing 211111, China (e-mail: \{czheng; s.liu; huangym; lxyang\}@seu.edu.cn).}

\thanks{W. Zhang is with School of Electrical Engineering and Telecommunications, The University of New South Wales, Sydney, NSW 2052, Australia, and also with the Purple Mountain Laboratories, Nanjing 211111, China (e-mail: w.zhang@unsw.edu.au).}

\thanks{Copyright (c) 20xx IEEE. Personal use of this material is permitted. However, permission to use this material for any other purposes must be obtained from the IEEE by sending a request to pubs-permissions@ieee.org.}

}

% The paper headers
\markboth{IEEE INTERNET OF THINGS JOURNAL,~Vol.~XX, No.~X, XXX~XXXX}
{Zheng \MakeLowercase{\textit{et al.}}: UNSUPERVISED RECURRENT FEDERATED LEARNING FOR EDGE POPULARITY PREDICTION IN PRIVACY-PRESERVING MEC NETWORKS
}

\maketitle
\begin{abstract}

Nowadays wireless communication is rapidly reshaping entire industry sectors. In particular, mobile edge computing (MEC) as an enabling technology for industrial Internet of things (IIoT) brings powerful computing/storage infrastructure closer to the mobile terminals and, thereby, significantly lowers the response latency. To reap the benefit of proactive caching at the network edge, precise knowledge on the popularity pattern among the end devices is essential. However, (i) the spatiotemporal variability of content popularity, (ii) the data deficiency in privacy-preserving system, (iii) the costly manual labels in supervised learning as well as (iv) the not independent and identically distributed (non-i.i.d.) user behaviors pose tough challenges to the acquisition and prediction of content popularities. In this article, we propose an unsupervised and privacy-preserving popularity prediction framework for MEC-enabled IIoT to achieve a high popularity prediction accuracy while addressing the challenges. Specifically, the concepts of local and global popularities are introduced and the time-varying popularity of each user is modelled as a model-free Markov chain. On this basis, we derive and validate the essential relationship between the local and global popularities and then propose an unsupervised recurrent federated learning (URFL) algorithm to predict the distributed popularity while achieving privacy preservation and unsupervised training. Moreover, a federated loss-weighted averaging (FedLWA) scheme for the parameter aggregation is further designed to alleviate the problem of non-i.i.d. user behaviors. Simulations indicate that the proposed framework can enhance the prediction accuracy in terms of a reduced root-mean-squared error by up to $60.5\%$--$68.7\%$ compared to other baseline methods, i.e., recommendation algorithms, centralized learning algorithms, and other distributed learning algorithms. Additionally, manual labeling and violation of users' data privacy are both avoided.
\end{abstract}


%\bigskip

\section{Introduction}
\label{sec1}

\IEEEPARstart{T}{he} emerging industrial Internet of things (IIoT), also colloquially known as Industry 4.0, interconnects isolated industrial assets by leveraging the growing ubiquity of wireless communication technologies. By harvesting the rich supply of data from various networked embedded sensors, this new paradigm promises the opportunity to revolutionize production and manufacturing \cite{WuH19, Yang20}. However, to process such an enormous amount of data and to handle the massive requests generated by ubiquitous wireless devices especially under the stringent requirements of reliability, latency, security and privacy, are incredibly challenging. Mobile edge computing (MEC), which co-locates storage and processing resources at the network edge, represents an effective framework to provision IIoT services \cite{Sisi18} and to mitigate the surging traffic burden of the data centers \cite{Miao18,Zhao20}. Dense deployment of edge nodes (ENs), i.e., radio access points or micro base stations, allows proximal and immediate access to the IIoT services. In an information-centric networking, proactive edge caching (EC) is considered a cost-effective approach to address the backhaul bottleneck problem \cite{Zhong20} and to reduce the content retrieval/handover latency \cite{Chen20}.

The explorations of optimal EC policies in MEC-enabled IIoT networks have been investigated in many previous studies \cite{Zahed20,GuS20,LiQ20,ZhangR20,DaiY20,ZhuH19}. However, many of relevant works, i.e., \cite{Zahed20,GuS20,LiQ20}, assume that the content popularity can be {\em a priori} given and remains constant during the services, which is actually inconsistent with the reality. To be more realistic, some works \cite{ZhangR20,DaiY20,ZhuH19} have considered the unknown popularity and explored the end-to-end learning approach to learn the EC policy directly from the request data so as to avoid the popularity prediction. Although circumventing popularity prediction by introducing the end-to end machine learning is indeed a research direction of the EC policy optimization, the popularity reflects the inherent pattern of user interests and directly determines the generation of content requests and thus, plays the most direct and decisive role in the EC policy. Significant improvements of EC performance provided by the popularity prediction have been demonstrated in literatures \cite{Zeng18,Zheng20,ZhengC21}. Therefore, in this paper, we focus on the investigation of popularity prediction. Generally, content popularity depends on the user interest, which is complicated and spatiotemporal varying and, thus, is unavailable in advance no matter which caching policy is applied \cite{Qian19}. When considering the spatiotemporal varying characteristic of content popularities in the real world, the performance of EC policies is largely determined by the selection mechanism of popular data that is worthwhile caching from a massive deluge of data traffic, which in turn relies on the accuracy of popularity prediction \cite{Zeng18, Zheng20, Liu2022}.

The potential of the popularity prediction has attracted the attentions of researchers and many progresses have been achieved in relevant works, e.g., \cite{Jiang19}--\cite{Kair20}. However, there are still open challenges for the popularity prediction. 1) {\em Spatiotemporal variability}: Due to the complex and changeable subjective attributes of end users/devices such as the subjective interests of human and the intrinsic task characteristics of devices, the content popularity is spatiotemporal varying and pose tough challenges to its prediction accuracy. 2) {\em Privacy preservation}: In many scenarios such as the healthcare and automotive-related industries, the terminal data is private and needs to be protected from external access. Thus, privacy requirements prevent the data sharing among devices and center servers, which leads to the data deficiency for the data-driven centralized popularity prediction methods. 3) {\em Costly manual labeling}: Caused by the unobservability of popularity in the realistic environment, labelling popularities in manual for the popularity prediction is costly and challenging, which brings a technical bottleneck to many prediction approaches based on supervised learning. 4) {\em Not independent and identically distributed (non-i.i.d.) behaviors}: Due to the subjectivity of user interests, user behaviors are non-i.i.d., which can violate the assumption in machine learning that datas are independent and identically distributed and sequentially causes many issues, i.e., feature distribution skew, concept drift, quantity skew, etc., for learning-based prediction methods.

In the light of the above observations, the objective of this study is to design a distributed deep learning algorithm to predict the dynamic content popularities in a MEC-enabled IIoT system while preserving the data privacy of end devices and circumventing the costly manual labeling. To explore the insightful relations between the local and global popularities is another important consideration in this paper. To this end, the mathematical derivation and simulation validation of the relations among the popularities in system are provided, and a novel unsupervised recurrent federated learning (URFL) algorithm with a novel federated loss-weighted averaging (FedLWA) parameter aggregation scheme are also proposed. The main technical contributions of this work are summarized as follows.

\begin{itemize}
\item We respectively introduce the time-varying local and global popularities in the local user and MEC server sides to make the MEC system more closely aligned with reality. Furthermore, we derive and validate the mathematical relationship between the local and global popularities, and reveal the fundamental difficulty in inferring the global popularity under the privacy-preserving constraint.
	
\item We design the learning node architecture in the FL framework by embedding the AE module, which realizes unsupervised learning without costly manual labeling. To effectively extract the underlying temporal information in the historical requests, long short-term memory (LSTM) cells are adopted in the AE module.

\item We propose a novel URFL algorithm on the basis of the FL architecture to perform offline training and online realtime prediction of the distributed popularities. The proposed URFL algorithm breaks the consistency requirement on model inputs between local and global sides in the typical FL framework and realizes the distributed training and prediction without any external access to the historical requests of local users except themselves, so as to better preserve user privacy.

\item On the basis of the proposed URFL algorithm, we further design a FedLWA scheme for the parameter aggregation to alleviate the problem of non-i.i.d. user behaviors considered in the investigated scenario and, thereby, further reduce the prediction error of popularities.
\end{itemize}

The rest of this paper is organized as follow. Section~\ref{Relwork} reviews the related works. The system model is established in Section~\ref{sec2}. Then, Section~\ref{sec3} introduces the problem formulation and the proposed scheme. In Section~\ref{sec4}, simulation results and discussions are provided. This article is concluded in Section~\ref{sec5}.

\section{Related Works} \label{Relwork}

\subsection{Popularity Prediction}
The complex and varied user interest/need poses enormous challenges for accurately predicting the dynamic popularity. To this end, many different algorithms have been proposed in the literature to predict the dynamic popularity over the recent years. Some inspiring examples include the time-series prediction method \cite{Jiang19, Zheng20}, the social-driven prediction method \cite{XuJ15, He17}, and the statistics-based prediction method \cite{Trzc17, Mehr19}. Jiang~et~al. \cite{Jiang19} proposed an online content popularity prediction algorithm by exploiting the content features and user preferences, where the user preferences were learned offline from the historically requested information. In \cite{Zheng20}, an auto-encoder (AE) neural network was combined with the long short-term memory module to predict the popularity by extracting time-series features from the historical requests of users. Nevertheless, the time-sequence prediction approaches in \cite{Jiang19,Zheng20} rely heavily on the privacy of users such as historical requests to improve the prediction accuracy, which is intolerable for those privacy-sensitive users and not appropriate for the privacy-preserving systems. Xu~et~al. \cite{XuJ15} explored the dynamically changing and evolving propagation patterns of videos in social media and the content popularity could be forecasted in a timely fashion. In \cite{He17}, the social relationships among a small number of users were explored to bridge the gap between prediction accuracy and small population. A social-driven propagation dynamics model was proposed therein to improve the popularity prediction accuracy. However, as the hidden features can only be extracted from massive amounts of social information accumulated over a long-term time, the timeliness and privacy concerns of the social-driven prediction methods are questioned. Statistical prediction methods based on regression analysis have also been examined. For instance, Trzcin\'ski~et~al. \cite{Trzc17} used support vector regression based on Gaussian radial basis functions to predict the online video popularity. Similarly, a Bayesian hierarchical probabilistic model was designed \cite{Mehr19} to regress the content popularity in an EC network. While the existing statistical approaches show potentials in achieving accurate and stable prediction, they are still far from practical use. For instance, the selection of probabilistic models in statistics-based methods \cite{Trzc17, Mehr19} has critical impacts on the prediction performance, but the selection criteria are unclear and impossible to be {\em a priori} given in the real world. Most importantly, they inevitably raise privacy issues due to the fact that statistics-based methods require access to the historical request log data of users for popularity prediction.

\subsection{Federated Learning}
As a matter of fact, private data leakage vulnerabilities in IIoT systems, especially in the healthcare and automotive-related industries, can lead to catastrophic consequences such as endangering user safety and causing severe property loss for data providers \cite{LuY20}. The resultant privacy preservation constraint makes the dynamic popularity prediction in EC even trickier. Recently, the disruptive blockchain technique \cite{Sharma18} shows superiority in enhancing data security and privacy preservation due to its anonymity, inherent decentralization, and trust properties. Nevertheless, the blockchain technique is essentially a distributed database of records and it has no interface for user behavior analysis  \cite{Xiong20}. We argue that the knotty problem of privacy-preserving popularity prediction can be tacked by leveraging the recent advances of distributed deep learning, particularly the federated learning (FL) \cite{LiuG21}. FL has emerged as a distributed artificially intelligence (AI) approach, by coordinating multiple devices to perform AI training without sharing raw data for privacy enhancement \cite{Nguy21}.

FL incorporating deep neural networks (DNNs), which combines the capabilities of DNNs in extracting features from input data and the advantages of FL in distributed training and privacy preservation, and has become one of the main paradigms of FL \cite{ZhouH21}. Therein, convolutional neural networks (CNNs) and recurrent neural networks (RNNs) are two important types used for the incorporation with FL
%Convolutional neural networks (CNNs) and recurrent neural networks (RNNs), as two typical DNNs structures, have been widely used in image processing and sequence data processing respectively.
Literature \cite{OhS20} investigates the image classification problem and proposes a communication-efficient and privacy-preserving distributed machine learning framework based on the FL cooperating with CNNs. The superior classification accuracy shown in \cite{OhS20} demonstrates the strong ability of the FL cooperating with CNNs in image feature extraction while preserving privacy. However, the architecture of CNNs is not appropriate for the feature extraction of sequence data which is rather important to wireless communication systems. RNNs, as an efficient architecture of sequence data processing, cooperating with FL is viewed as a promising framework for privacy-preserving data processing in wireless communication systems. For instance, Liu et al. \cite{LiuY20} provide a FL-based gated recurrent unit neural network for traffic flow prediction while providing reliable data privacy preservation. Although FL integrating with DNNs has been widely investigated, many existing works,\cite{OhS20,LiuY20}, adopt the supervised learning with costly manual labels which poses significant challenges to their practical applications, especially to practical popularity prediction. The spatiotemporal variability and unobservability of the content popularity lead to the difficulty in manually obtaining the popularity labels. Therefore, we extending FL to an unsupervised paradigm in this paper to address the challenges on the manual labelling of popularities. In this paper, we proposed an unsupervised FL incorporating RNNs to effectively predict popularities from sequences of historical requests without labels while preserving user privacy. Moreover, we further design a parameter aggregation to alleviate the non-i.i.d. problem which is an open challenge in the research field of FL \cite{Kair20}.

%Actually, many open challenges regarding accuracy and local training feasibility in unsupervised FL remain to be addressed \cite{Kair20}.
%\cite{Edge Device Identification Based on Federated Learning and Network Traffic Feature Engineering} investigates the access control and management of the suspicious devices in internet of things (IoT) systems and develops a security access mechanism based on the FL collaborating convolutional neural networks (CNNs)

\subsection{FL-based Popularity Prediction}
The MEC framework enables FL in the wireless communication networks with the supply of abundant and closer computing/caching resources. A comprehensive survey of FL from the perspective of fundamentals, challenges, solutions, and applications in MEC networks can be found in \cite{Lim20}. %Nguyen~et~al. further investigated utilizing blockchain to further improve the security of FL implementation \cite{Nguy21}.
Therein, FL-based privacy-preserving popularity prediction in MEC networks has been explored in many works, i.e.,\cite{Sapu19,YuZ18,CuiL20,QiK20,YuZ21}. Nevertheless, many challenges still remain to be addressed. In a recent work \cite{Sapu19}, the center server is prohibited from snooping on users' private data, while only the local MEC server is permitted to collect and learn from the historical requests of users. This scheme relies on authorization management and is susceptible to unauthorized access provided by the unreliable network operator or gained by malicious cracking. In addition, the deep learning method in \cite{Sapu19} requires manual labeling in advance, which unfortunately is costly and infeasible in real implementation. The authors in \cite{YuZ18} proposed an FL-based method to realize the privacy-preserving EC. On the basis of literature \cite{YuZ18}, Cui~et~al. \cite{CuiL20} utilized blockchain to further improve the security of FL implementation. %In \cite{CuiL20}, the $K$-means clustering method is applied to calculate similarities between users and contents according to potential features extracted by neural network from historical requests of users, and then MEC servers cache the $K$-th most similar contents based on the similarities.
However, the EC policies considered in \cite{YuZ18,CuiL20} are both built on the similarities between users and contents, which is calculated by potential features extracted from historical requests of users. The similarity calculated in these two literatures is just a rough estimation of popular contents rather than the actual popularity. Thus, the content popularity which represents the accurate requested probability of each content has not been predicted in \cite{YuZ18,CuiL20}. Moreover, the relationship between the client-side popularity and server-side popularity has not been explored in \cite{YuZ18,CuiL20}. In \cite{QiK20}, the content popularity has been predicted while the popularity relationship between the client side and server side is preliminarily explored. Nonetheless, the explorations of the relationship between client-side and server-side popularities were sketchy and empirical in \cite{QiK20}, which was reflected in the thoughtlessness of user request arrival rate as well as the absence of any verification for the given relationship. In this paper, we introduce the concept of local popularity at user side and global popularity at MEC server side respectively, and further derive and validate the mathematical relationship between these two popularity types.

Yu~et~al. \cite{YuZ21} considered the mobility of users and proposed a mobility-aware FL method to predict the popularity while preserving user privacy. Nevertheless, the temporal variability of the popularity caused by the time-varying user interests has not been explored in \cite{YuZ21}. Furthermore, in \cite{YuZ21}, sampling from the real popularity distribution at the MEC server side is required to generate the input of the prediction model during the local training phase. However, the real popularity distribution at the MEC server side depends on the subjective interests of users within the entire service area and thus is extremly hard to obtain {\em a priori}. In our study, we consider the local and global popularities both time-varying and unavailable so as to be more closely aligned with reality. In addition, we can further observe from \cite{YuZ18,CuiL20,QiK20,YuZ21} that local users are supposed to share or upload their request history when making online predictions, so as to generate the input of the prediction model. Indeed, this kind of sharing or upload violates the privacy-preserving requirement. The fundamental reason why these works need to collect users' request history for the global prediction can be attributed to the consistency requirement in the typical FL framework, which demands that the model structures and model inputs should be exactly the same between the local and global sides \cite{Kair20}. To mitigate the risk of privacy leakage caused by the collection of user request history, we design the input structure of the local and global models to break the consistency requirement on model inputs, and then realize the distributed training and prediction without any external access to user's historical requests except itself.
%In addition, the privacy preservation is incomplete under the proposed FL-frawework in \cite{QiK20} due to the fact that each user therein is required to upload its content preference and historical requests to the MEC server.

\begin{figure}[!b]
\centering
\includegraphics[width=0.375\textwidth]{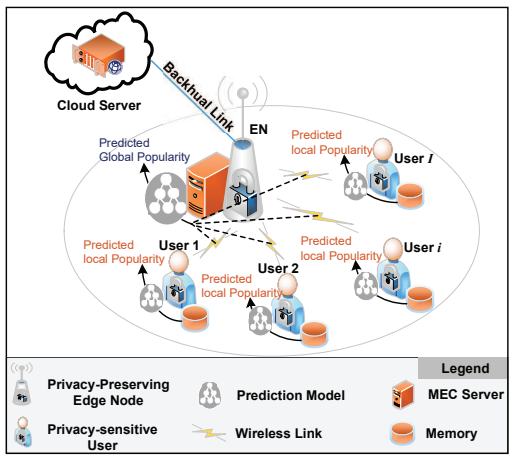}
\caption{Hierarchical architecture of the privacy-preserving and MEC-enabled network under investigation.}
\label{fig1}
\end{figure}

\section{SYSTEM MODEL}
\label{sec2}

This section provides brief descriptions of the system model and the underlying concepts used in this work. A hierarchical wireless network of shared caches is considered for smart industry applications, and the network is supposed to be able to provide a secure and trustworthy content service. As illustrated in Fig.~\ref{fig1}, a cloud server is deployed by the service providers to store contents for the consumers, i.e., the mobile end-users and/or IIoTs devices. One MEC server, which can be the general-purpose computer or server, is placed on the edge node between the cloud and the end sides. The edge node can directly provide the content services supported by the caching capability of the MEC server. Moreover, the prediction models for popularity prediction can be placed on the user equipments (UEs) and the edge node benefited from the rapidly evolving computing capabilities of the UEs and the MEC server. In the considered scenario, we assume that a total of $I$ privacy-sensitive users are directly connected to an edge node within a certain small cell, where the edge node must obey some privacy-preserving mechanism to preserve user privacy. \footnote{Note that the security/privacy analysis by defining threat models or hacker attacks against the privacy-preserving mechanism is a meaningful but challenging research direction. (c.f., e.g., \cite{LiuY19,MaZ20,Ferr18,LiuY21}). However, this paper focuses on popularity prediction in a privacy-preserving wireless MEC network. To avoid further complicating the problem under investigation, the backdoor problems are left for our future work.}

At the beginning of a time slot, a user will send a content request to the edge node if a certain content file cannot be found in its local cache. Though much closer to the users, the edge nodes have limited caching and computing capabilities compared to the cloud server. As such, the edge nodes can only store some selective --- usually most popular contents. Whenever the requested contents are not located in the edge nodes, the request will be further forwarded to the cloud via the backhaul link. Additionally, definitions of key notations used in this paper are given in Table \ref{notations_def} for ease of reading.

%fetch the content from the cloud server

\subsection{Service Process}\label{ser_pro}

We assume that the content library contains $N$ files, which is denoted as a set ${{\cal F}} = \{ {{F}_1},{{ F}_2}, \ldots ,{{F}_N}\}$, and the cloud have a complete copy of all the files. Limited by the cache capacity, the MEC server can only cache ${M_0}$ files and an arbitrary UE-$i$ can cache ${M_i}$ files. Generally, ${M_0} \gg {M_i},\forall i \in {\cal I} = \{ 1,2, \ldots ,I\}$. We assume that a content file ${F^i}(t) \in \emptyset  \cup {{\cal F}}$ is requested by UE-$i$ at time slot $t$, where we have ${F^i}(t) \in \emptyset $ when UE-$i$ does not request any contents. Then, the request will be uploaded to the MEC server if ${F^i}(t)$ cannot be found in UE-$i$, which is represented as $F^{i}\left(t\right)\notin\mathcal{C}_{i}(t)$, where $\mathcal{C}_{i}(t)$ is the files set cached in UE-$i$ at time slot $t$. Conversely, if ${F^i}(t) \in \emptyset$ or $F^{i}\left(t\right)\in\mathcal{C}_{i}(t)$ is true, UE-$i$ will not upload this request information to the MEC server. MEC server will retrieve content for the received requests in its current cache $\mathcal{C}_{0}(t)$. If $F^{i}\left(t\right)\notin\mathcal{C}_{0}(t)$ is satisfied, the MEC server will further request the absent files from the cloud. Finally, the requested files of each user will be sent back from the MEC server. In addition, it is worth to note that the users will not upload any request information in time slot $t$ unless $F^{i}\left(t\right)\notin\mathcal{C}_{i}(t)$.

\begin{table*}[!t]
\centering
\caption{DESCRIPTIONS OF KEY NOTATIONS} \label{notations_def}
\begin{tabular}{cl}
\toprule[1pt]
    \textbf{Notation} & \textbf{Description} \\
    \hline\hline
              $t$ &  Index of time slot. \\
%    \hline
          ${\cal I} = \{ 1,2, \cdots ,I\}$ &   Set of UE indexs. \\
%    \hline
      ${{\cal F}} = \{ {{F}_1},{{ F}_2}, \cdots ,{{F}_N}\}$ &  Set of all contents. \\
%%    \hline
%      ${M_0}$, ${M_i}$  & Storage capacity of the MEC server and UE-$i$, respectively. \\
%%    \hline
%           $\mathcal{C}_{0}(t)$, $\mathcal{C}_{i}(t)$ &  Content set respectively cached in the MEC server and UE-$i$ at time $t$. \\
%    \hline
      ${F^i}(t)$ & Content request generated by UE-$i$ at time $t$. \\
%    \hline
      ${\bf P}^{i}\left(\alpha^{i}\left(t\right),t\right)$ & Local popularity on UE-$i$ at time $t$.\\
%     \hline
      $P_{n}^{i}\left(\alpha^{i}\left(t\right),t\right)$ & Probability that content $F_N$ is requested by UE-$i$ at time $t$.\\
%     \hline
	  ${{\bf{P}}^{\rm{G}}}(t)$ & Global popularity on the MEC server at time $t$.\\
%    \hline
	  $P_n^{\rm{G}}(t)$ & Probability that content $F_n$ is requested within the service area at time $t$.\\
%    \hline
	  ${\bf {\hat P}}^{i}\left(\alpha^{i}\left(t\right),t\right)$ & Prediction of the local popularity on UE-$i$ at time $t$.\\
%    \hline
	  ${{\bf{\hat P}}^{\rm{G}}}(t)$ & Prediction of the global popularity on MEC server at time $t$.\\
%    \hline
      $\alpha^{i}\left(t\right)$&  Probability distribution parameter of ${\bf P}^{i}\left(\alpha^{i}\left(t\right),t\right)$ at time $t$. \\
%    \hline
      $\lambda_{i}\left(t\right)$ & Content request arrival rate of UE-$i$ at time $t$.\\
%    \hline
      ${{\cal G}_i} = \{ \alpha _{g}^i|g = 1,2, \cdots ,{G_i}\}$ & Parameters set that $\alpha^{i}\left(t\right)$ evolves over time.\\
%    \hline
      $\mathbf{P}_{i}=\left\{P_{g_{l}g_{k}}^{i}\right\}_{g_{l},g_{k}=0}^{G_{i}}$ & Transition probability matrix of $\alpha^{i}\left(t\right)$.\\
%    \hline
      ${{\bf{R}}^i}(t) = [{{ F}^i}(t - H), \cdots ,{{ F}^i}(t)]$                 &  Extractor of UE-$i$ to extract its historical request information.\\
%    \hline
      ${\bf{R}}^{\rm{G}}(t) = \left\{{{ F}^i}(t)\right\}_{i = 1}^I$         &  Request information received by the MEC server at time $t$. \\
%    \hline
      $H$   &  Observation window length  of the extractor.\\
%    \hline
      ${f^{{\Theta ^i}}}( \cdot )$         &  Local popularity prediction model inside UE-$i$. \\
%    \hline
	  $f^{{\Theta ^{\rm{G}}}}( \cdot )$         &  Global popularity prediction model at the MEC server side. \\
%    \hline
      ${{\Theta ^{\rm{G}}}}$, ${{\Theta ^{\rm{i}}}}$ & Parameters set of  global and local popularity prediction model respectively. \\
%    \hline
%      $x^t$, $y^t$, $C^{t}$ & Input, output, and memory state of the LSTM cell, respectively, at time slot $t$. \\
%%    \hline
%      $\sigma$ , $o$  & Control gate and output gate of the LSTM cell, respectively. \\
%%    \hline
%	  $W$ & Dependency evaluations of the weight parameters in the LSTM cell \\
%%    \hline
%	  $b$ & Offset parameters in the LSTM cell \\
%    \hline
	  $\mathbf{z}^{l_{\textrm{e}}}\left(t\right)$ & Output of the encoding function in the $l_{\textrm{e}}$-th layer at time $t$ \\
%    \hline
	  $\mathbf{\hat{z}}^{l_{\textrm{d}}}\left(t\right)$ & Output of the decoding function in the $l_{\textrm{d}}$-th layer at time $t$ \\
%    \hline
	  $L_{\rm{e}}$, $L_{\rm{d}}$ & Hidden layer number of the encoder and decoder, respectively.\\
%    \hline
      ${{\cal D}^i} = \{ {{ F}^i}(t)|t \in {\mathbb Z}_{0+} \}$ & Historical request data of UE-$i$. \\
%    \hline
%      ${{\cal T}^i}$ & Training dataset of the local popularity prediction model in UE-$i$. \\
%    \hline
      $T$ & Number of local training at every communication round. \\
%    \hline
      $L({\Theta ^i})$ & Training loss function of the local popularity prediction model in UE-$i$. \\
%    \hline
      ${\Theta ^{\rm{L}}}$ & Stack of parameter sets uploaded by all users \\
%    \hline
      ${\Theta ^{\rm{AE}}}$ & Updated parameters set aggregated from ${\Theta ^{\rm{L}}}$.\\
%    \hline
      ${\Theta _{\rm{E}}^i}$, ${\Theta _{\rm{D}}^i}$ & Parameters set of encoder and decoder on UE-$i$, respectively.\\
%    \hline
      ${L_{\rm{Avg}}}\left( {{\Theta ^i}} \right)$  & Average training loss of UE-$i$ at each communication round.\\
%    \hline
      $\gamma_{i}$  & Aggregation weight for the model parameters of UE-$i$ in the FedLWA scheme.\\
%    \hline
\bottomrule[1pt]
\end{tabular}
\end{table*}
\subsection{Local and Global Popularity}\label{lgpopu}

As mentioned above, the content popularities on the local user side and the MEC server side are respectively termed \emph{local popularity} and \emph{global popularity}. Regarding the local popularity, we assume that the content popularity of each user in each time slot $t$ follows a Zipf distribution which has been wildly adopted in related works \cite{Sade17,cui16,YangL20}. Moreover, the time-varying nature of the popularity is taken into account in this paper. For an arbitrary UE-$i$, the probability of demanding the $n$-th file at time slot $t$ is
\begin{equation} \label{e1}
P_{n}^{i}(\alpha^{i}(t),t)=\left(n^{\alpha^{i}(t)}\sum\limits _{l=1}^{N}l^{-\alpha^{i}(t)}\right)^{-1},
\end{equation}
where the $N$ files have been assigned with a descending ordering of popularity in each time slot $t$. The distribution parameter ${\alpha ^i}(t)$ evolves over time. As such, the content popularity of UE-$i$ at $t$ can be denoted as ${\bf P}^{i}\left(\alpha^{i}\left(t\right),t\right)=\left\{ P_{n}^{i}\left(\alpha^{i}\left(t\right),t\right)\right\} _{n=1}^{N}$. It should be noticed that the Zipf distribution is assumed for the convenience of discussion, and generalization to any other probability distribution model is straightforward.

\begin{figure}[!t]
\centering
\includegraphics[width=0.48\textwidth]{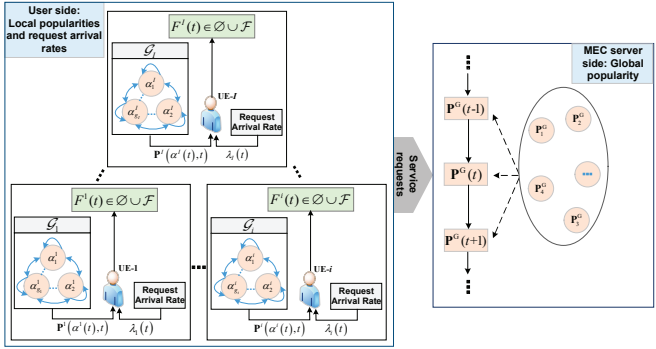}
\caption{Dynamic models of local and global popularity.}
\label{fig2}
\end{figure}

As depicted in Fig. \ref{fig2}, for user $\forall i \in {\cal I}$, we model the dynamics of ${\alpha ^i}(t)$ using a model-free Markov chain with the states $\left| {{{\cal G}_i}} \right|$ recorded in the set ${{\cal G}_i} = \{ \alpha _g^i|g = 1,2, \ldots ,{G_i}\}$. Consequently, the dynamics of ${\alpha ^i}(t)$ can be defined as
\begin{equation} \label{e1.11}
P\left\{ \alpha^{i}\left(t+1\right)=\alpha_{g}^{i}\right\} =P_{\alpha^{i}\left(t\right)\rightarrow\alpha_{g}^{i}}^{\alpha^{i}\left(t\right)},\forall\alpha_{g}^{i}\in\mathcal{G}_{i},
\end{equation}
where $P_{\alpha^{i}\left(t\right)\rightarrow\alpha_{g}^{i}}^{\alpha^{i}\left(t\right)}$ denotes the transition probability of $\alpha^{i}\left(t\right)$ transits to $\forall\alpha_{g}^{i}\in\mathcal{G}_{i}$. It is worth mentioning that, neither the parameter sets nor the transition probabilities are unknown to the model-free Markov chain due to the diversity and complexity of users' subjective interests \cite{Zheng20}. The difference among the users are captured by the set ${{\cal G}_i}$ as well as the potential state transition probabilities. Besides, user behaviors are assumed to be non-i.i.d. in the system model.

At the MEC server side, the global popularity at time slot $t$ can be represented as ${{\bf{P}}^{\rm{G}}}(t) = \left\{P_n^{\rm{G}}(t)\right\}_{n = 1}^N$, where $P_n^{\rm{G}}(t)$ is the probability that content $n$ is requested within the service area. Apparently, the global popularity depends on all the local popularities within the service area, and the MEC server as a service provider can significantly improve its caching efficiency under the guidance of accurate knowledge of the global popularity. However, as will be elaborated in Section~\ref{sec3}, the prediction of the global popularity is much more complicated than that of the local popularity due to the different behavior patterns of different users as well as the privacy-preserving constraint.

\subsection{Content Request Model}

At time slot $t$, a certain UE-$i$ requests a file $F^{i}(t)\in\emptyset \cup\mathcal{F}$, where the probability of $F^{i}(t)\in\emptyset$ follows its corresponding current request arrival rate, denoted as $\lambda_{i}\left(t\right)$. If $F^{i}(t)\in \mathcal{F}$, UE-$i$ make a request, which satisfies the present probability distribution denoted as $F^{i}\left(t\right)\sim {\rm{Zipf}}\left(\alpha^{i}\left(t\right)\right)$. According to the above description, the content request model of an arbitrary UE-$i\in\mathcal{I}$ at time slot $t$ can be expressed as
\begin{equation} \label{e1.1}
F^{i}\left(t\right)\in\begin{cases}
\emptyset, & P\left\{ F^{i}\left(t\right)\in\emptyset\right\} =1-\lambda_{i}\left(t\right)\\
\mathcal{F}, & P\left\{ F^{i}\left(t\right)=F_{n}\right\} =\lambda_{i}\left(t\right)\cdot P_{n}^{i}\left(\alpha^{i}\left(t\right),t\right),
\end{cases}
\end{equation}
where $F_{n}\in\mathcal{F}$ and $\alpha^{i}\left(t\right)\in\mathcal{G}_{i}$. Similar to ${\alpha ^i}\left(t\right)$, parameter $\lambda_{i}\left(t\right)$ also reflects the individual characteristics of user $i$ and should be protected from being accessed by others.

\subsection{Privacy-preserving Mechanism}\label{Prp_mec}

In the real world, privacy-sensitive users usually concern about the leakage of their private data such as location information, historical contents/services request, personal bank or social account information. In this paper, we readily observed from the service process that the private information of users involved in the problem under investigation is mainly the historical contents request data, and the historical request database of each user $\mathcal{D}^{i}=\left\{ \left.F^{i}\left(t\right)\right|t=1,2,\cdots\right\}$ is stored only in their own UEs and is inaccessible to outsiders. Moreover, as stated in some data privacy legislations such as the European Commission's General Data Protection Regulation (GDPR) \cite{GDPR16}, users have the right to require the responsible party to delete the individual data records about them. Thus, to respond with the implementation of GDPR, the burn-after-read principle as the privacy-preserving mechanism is implemented in the MEC servers, i.e., the request information from users must be immediately deleted from the memory of the MEC server once the contents have been scheduled. The MEC servers are not allowed to hold any historical information of any users.

%we investigate the problem of content popularity prediction in a privacy-preserving MEC system, which is critical for the subsequent optimization of EC policy.

\begin{figure*}[!t]
\centering
\includegraphics[width=0.62\textwidth]{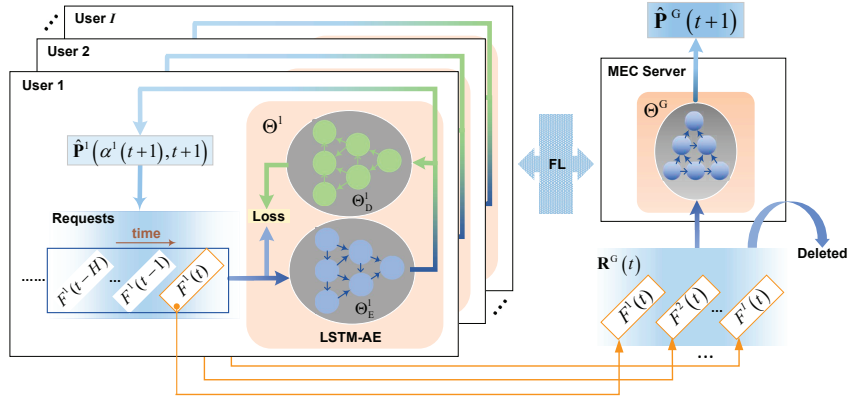}
\caption{Network architecture of the URFL algorithm.}
\label{fig3}
\end{figure*}

\section{URFL FOR EDGE POPULARITY PREDICTION}
\label{sec3}

\subsection{Problem Formulation}

In the MEC-based IIoT system under investigation, both the user and the server sides participate in the popularity prediction process. Given a popularity ${{\bf{P}}^i}({\alpha ^i}(t),t)$ at time slot $t$, UE-$i$ generate a content request denoted as ${{ F}^i}(t){|_{{{\bf{P}}^i}({\alpha ^i}(t),t)}}$, which is simplified as ${{F}^i}(t)$ in the sequel for the convenience of discussion. Based on the request history saved in its local memory, the local popularity of user $i$ can be predicted by
\begin{equation} \label{e11}
{{{\bf{\hat P}}}^i}({\alpha ^i}(t + 1),t + 1) = {f^{{\Theta ^i}}}({{\bf{R}}^i}(t)),
\end{equation}
where ${f^{{\Theta ^i}}}( \cdot )$ is a predictive model inside UE-$i$ and ${\Theta ^i}$ denotes the collection of trainable parameters therein. ${\bf R}^{i}(t)=\left[F^{i}(t-H),F^{i}(t-H+1),\cdots,F^{i}(t)\right]$ is a extractor of UE-$i$ to extract its historical request information of continuous $H$ times before time $t$. $H$ is the observation window length of the extractor. On account of the structural features of AE, we can divide ${\Theta ^i}$ into encoder and decoder parameters sets, denoted as ${\Theta ^i} = \left\{\Theta _{\rm{E}}^i,\Theta _{\rm{D}}^i\right\}$. The implementation of ${f^{{\Theta ^i}}}( \cdot )$ will be detailed in Section~\ref{method}. %${{{\bf{\hat P}}}^i}({\alpha ^i}(t + 1),t + 1)$ is the prediction of ${{\bf{P}}^i}({\alpha ^i}(t + 1),t + 1)$.

At time $t$, to minimize the distributed popularity prediction errors of all the users at future time $t+1$, the mean-square error (MSE) metric is adopted and the underlying optimization problem of arbitrary UE-$i$ can be formulated as
\begin{subequations}\label{P_i}
    \begin{align}
    P_i:
    \underset{\Theta^{i}}{\textrm{min}} \quad
    &\frac{1}{N}\left\Vert {\bf P}^{i}(t+1)-{\bf \hat{P}}^{i}(t+1)\right\Vert _{2}^{2},\label{P_i_a}\\
    {\rm{s.t.}}\quad
    & \left\Vert \mathbf{R}^{i}\left(t\right)\right\Vert _{0}\leq H,\label{P_i_b}\\
    &\alpha^{i}(t-h)\in\mathcal{G}_{i},\forall h\in\left\{ 0,\cdots,H-1\right\},\label{P_i_c}\\
    &0\leq\lambda_{i}\left(t-h\right)\leq1,\forall h\in\left\{ 0,\cdots,H-1\right\},\label{P_i_d}\\
    &F^{i}(t-h)\in\mathbf{\mathcal{D}}^{i},\forall h\in\left\{ 0,\cdots,H-1\right\},\label{P_i_e}
    \end{align}
\end{subequations}
where ${\bf P}^{i}(\alpha^{i}(t+1),t+1)$ and ${\bf \hat{P}}^{i}(\alpha^{i}(t+1),t+1)$ are abbreviated as ${\bf P}^{i}(t+1)$ and ${\bf \hat{P}}^{i}(t+1)$, respectively. $\left\Vert \cdot\right\Vert$ and $\left\Vert \cdot\right\Vert_{2}$ respectively represent the $l_0$ and $l_2$ norm. Constraint \eqref{P_i_b} ensures the observation window length of the extractor $\mathbf{R}^{i}\left(t\right)$ at time $t$ not exceed $H$. $\alpha^{i}(t-h)$ and $\lambda_{i}\left(t-h\right)$ depend on the subject interests of user $i$ at time $t-h$. $F^{i}(t-h)$ is the component of $\mathbf{R}^{i}$ and extracted from the request database $\mathbf{\mathcal{D}}^{i}$. %Note that the problem \eqref{P_i} evolves over time and the optimal solution $\Theta^{i*}$ should be valid at any time $t$.
Note that ${\bf P}^{i}(\alpha^{i}(t+1),t+1)$, $\mathcal{G}_{i}$, $\alpha^{i}(t-h)$ and $\lambda_{i}\left(t-h\right)$ are all time-varying and unknown, which poses significant challenges to solve the problem \eqref{P_i}.

The MEC server makes the global prediction in a completely different way since there is no historical information of any users under the privacy-preserving constraint. The only data that the MEC server can provisionally acquire is ${\bf{R}}^{\rm{G}}(t) = \left\{{{ F}^i}(t)\right\}_{i = 1}^I$ at time slot $t$, which will be erased from the server before the next time slot. To further evaluate the difficulty in predicting the global popularity in such cases, we first give the following theorem which reveals the mathematical relationship between the local and global popularities.

\begin{theorem}\label{theorem1}
Given the local popularity $\left\{ {{{\bf{P}}^i}({\alpha ^i}(t),t)} \right\}_{i = 1}^I$ and the request-arrival rate of each user $\left\{ {{\lambda _i}(t)} \right\}_{i = 1}^I$ at time slot $t$. The global popularity ${{\bf{P}}^{\rm{G}}}(t)$ at the MEC server side is:
\begin{equation}\label{e12}
{{\bf{P}}^{\rm{G}}}(t) = \frac{{\sum\limits_{i = 1}^I {{\lambda _i}(t) \cdot {{\bf{P}}^i}\left( {{\alpha ^i}(t),t} \right)} }}{{\sum\limits_{i = 1}^I {{\lambda _i}(t)} }},
\end{equation}
\end{theorem}
\begin{IEEEproof}
The proof is presented in Appendix \ref{theorem1proof}
\end{IEEEproof}

The simulation validation of Theorem \ref{theorem1} can be found in Appendix \ref{theo1_simu}. According to Theorem \ref{theorem1}, we find that ${{\bf{P}}^{\rm{G}}}(t+1)$ cannot be obtained without any \emph{a priori} knowledge of $\left\{ {{{\bf{P}}^i}({\alpha ^i}(t),t)} \right\}_{i = 1}^N$ and $\left\{ {{\lambda _i}(t)} \right\}_{i = 1}^I$. Nonetheless, the local popularity and the request-arrival rate of each user both dynamically varies over time and also cannot be acquired in the privacy-preserving system. To address this challenge, a URFL algorithm is proposed in this work to predict the global popularity without violating UEs' data privacy. By employing the URFL algorithm, the global popularity in the next time slot $t+1$ is predicted by exploiting the newly arrived request at time slot $t$, i.e.,
\begin{equation} \label{e15}
{{{\bf{\hat P}}}^{\rm{G}}}(t + 1) = f^{{\Theta ^{\rm{G}}}}({\bf{R}}^{\rm{G}}(t)),
\end{equation}
where $f^{{\Theta ^{\rm{G}}}}( \cdot )$ represents the predictive function at the MEC server side in the proposed global model, and ${{\Theta ^{\rm{G}}}}$ is the parameters set. The implementation of $f^{{\Theta ^{\rm{G}}}}( \cdot )$ and ${{\Theta ^{\rm{G}}}}$ under the privacy-preserving mechanism will be detailed in Section~\ref{method}.

At the MEC server side, the underlying optimization problem at time $t$ can be formulated as
\begin{subequations}\label{P_G}
    \begin{align}
    P_{\rm G}:
    \underset{\Theta^{\textrm{G}}}{\textrm{min}} \quad
    &\frac{1}{N}\left\Vert {\bf P}^{\textrm{G}}(t+1)-{\bf \hat{P}}^{\textrm{G}}(t+1)\right\Vert _{2}^{2},\label{P_G_a}\\
    {\rm{s.t.}}\quad
    & \left\Vert \mathbf{R}^{\textrm{G}}\left(t\right)\right\Vert _{0}\leq I,\label{P_G_b}\\
    &\alpha^{i}(t)\in\mathcal{G}_{i},\forall i\in\mathcal{I},\label{P_G_c}\\
    &0\leq\lambda_{i}\left(t\right)\leq1,\forall i\in\mathcal{I},\label{P_G_d}\\
    &F^{i}(t)\in\mathbf{\mathcal{D}}^{i},\forall i\in\mathcal{I},\label{P_G_e}
    \end{align}
\end{subequations}
Similar with the problem $\eqref{P_i}$, the problem \eqref{P_G} also evolves over time and the optimal solution $\Theta^{\textrm{G}*}$ should be valid at any time $t$. However, ${\bf P}^{\textrm{G}}(t+1)$ is unknown while $\alpha^{i}(t),\lambda_{i}\left(t\right),\mathcal{G}_{i},\mathbf{\mathcal{D}}^{i},\forall i\in\mathcal{I}$ are unavailable to the MEC server due to the subjectively of user interests as well as the privacy-preserving requirements. Thus, it is also quite challenging to solve problem \eqref{P_G}.

\subsection{Methodology}\label{method}

This subsection presents the design of the URFL framework and its application to the privacy-preserving edge popularity prediction. The proposed architecture of URFL is illustrated in Fig.~\ref{fig3}. Concretely, in individual UE, a moderate-scale recurrent neural network (RNN) is prepared and trained on the local request history alone. The RNN in each UE is designed as an AE, with each neuron being an LSTM cell to capture the contextual information hidden in the input data \cite{RenL20, Huss20}. Since the MSE loss of an AE is directly obtained by comparing the input and output, the troublesome training data labeling is circumvented. Then, the collaborative prediction is performed under an FL framework, where the distributed UEs periodically exchange their diverse model parameters, rather than the raw private data, with the MEC server to collectively train a global model. Note that the MEC server only needs to deploy an encoder module. We next elaborate on each key element in the designed framework.

%preserving clients' private data from being exposed to adversaries

%In the proposed scheme, a pair of LSTM encoder-decoder are deployed to all the UEs in advance

\subsubsection{LSTM-AE Hierarchy}

\begin{figure}[!b]
\centering
\includegraphics[width=0.46\textwidth,trim=0 2 1 3,clip]{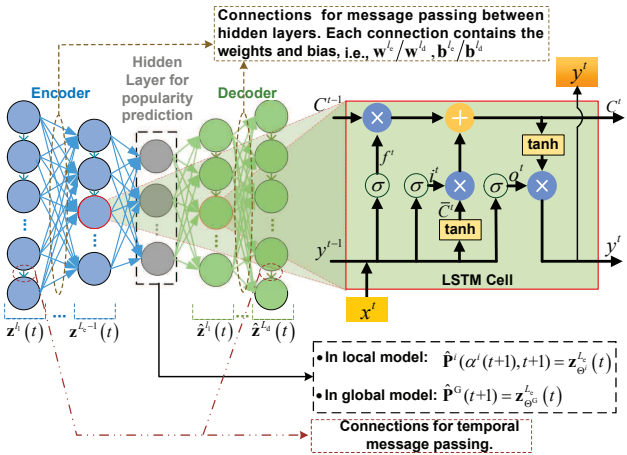}
\caption{Proposed LSTM-AE architecture.}
\label{figLSTMcell}
\end{figure}

RNNs perform hierarchical processing on complicated temporal tasks and, as such, it is capable of naturally capturing the underlying temporal dependencies in time series. In this work, we use a special type of RNN building block, i.e., LSTM cells to explore the evolving short-term dependencies within the long historical request sequences ${\bf{R}}^{i}(t)$ and ${\bf{R}}^{\rm{G}}(t)$ \cite{ZuoY20} and, in turn, predict the edge popularity more effectively. LSTM-based RNNs address the issue of vanishing gradients by integrating gating functions into their state dynamics \cite{Karim19}. As mentioned above, each UE has a built-in pair of LSTM encoder-decoder, whose hierarchical structure is given in Fig.~\ref{figLSTMcell}. Each neuron in the hierarchy is an LSTM cell, and each subsequent layer receives the hidden state of the previous layer as input time series. The auto-encoder architecture is created by symmetrically stacking the LSTM layers at the input and output sides, which respectively constitute the encoder and decoder. The iterative formula of message passing in one LSTM cell is as follows:
\begin{subequations}
\begin{equation}
f^{t}=\sigma\left(W_{f}\cdot\left[y^{t-1},x^{t}\right]+b_{f}\right),
\end{equation}
\begin{equation}
i^{t}=\sigma\left(W_{i}\cdot\left[y^{t-1},x^{t}\right]+b_{i}\right),
\end{equation}
\begin{equation}
\overline{C}^{t}=\tanh\left(W_{C}\cdot\left[y^{t-1},x^{t}\right]+b_{C}\right),
\end{equation}
\begin{equation}
o^{t}=\sigma\left(W_{o}\cdot\left[y^{t-1},x^{t}\right]+b_{o}\right),
\end{equation}
\begin{equation}
C^{t}=f^{t}*C^{t-1}+i^{t}*\overline{C}^{t},
\end{equation}
\begin{equation}
y^{t}=o^{t}*\tanh\left(C^{t}\right),
\end{equation}
\end{subequations}
where $x^t$, $y^t$, and $C^{t}$ respectively denote the input, output, and the memory state of the LSTM cell at time slot $t$. $\sigma$ is the control gate, which is typically a {\em Sigmoid} function. $f$ represents the output of the forgetting gate. $i$ and $o$ denote the output of the input and output gates, respectively. $W$ evaluates the dependencies of the weight parameters and $b$ denotes the offset parameter. By feeding the historical requests to the RNN network composed of these LSTM layers, we capture the features hidden in the input sequences. %Note that the dimension of the hidden layers shown in Fig.~\ref{figLSTMcell} is determined by the number of content files $N$.

\begin{figure}[!t]
\centering
\includegraphics[width=0.48\textwidth]{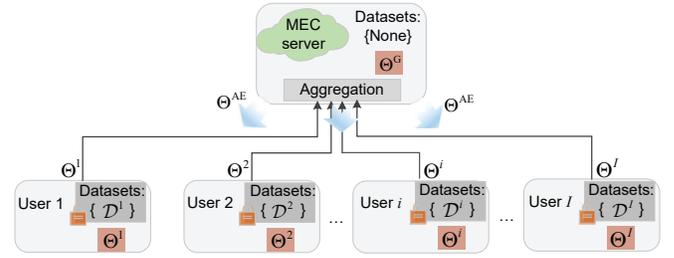}
\caption{Parameters passing during the training of the URFL algorithm.}
\label{param_pass_URFL}
\end{figure}

Then, the output of the encoding function in the $l_{\textrm{e}}$-th ($\forall l_{\textrm{e}}\in\left\{ 1,2,\ldots,L_{\textrm{e}}\right\}$) layer at time $t$ can be expressed as
\begin{equation}
\mathbf{z}^{l_{\textrm{e}}}\left(t\right)=\varphi\left(\mathbf{w}^{l_{\textrm{e}}}\cdot\left[\mathbf{z}^{l_{\textrm{e}}-1}\left(t\right),\mathbf{z}^{l_{\textrm{e}}}\left(t-1\right),\mathbf{C}^{l_{\textrm{e}}}\left(t-1\right)\right]+\mathbf{b}^{l_{\textrm{e}}}\right)
%\mathbf{z}^{l_{\textrm{e}}}=E\left(\mathbf{z}^{l_{\textrm{e}}-1}\right)=\varphi\left(\mathbf{w}^{l_{\textrm{e}}}\cdot\mathbf{z}^{l_{\textrm{e}}-1}+\mathbf{b}^{l_{\textrm{e}}}\right),
\end{equation}
where $\mathbf{w}^{l_{\textrm{e}}}$ and $\mathbf{b}^{l_{\textrm{e}}}$ respectively denotes the weight and implicit bias parameters of layer $l_{\textrm{e}}$ of the encoder. $\mathbf{z}^{l_{\textrm{e}}-1}\left(t\right)$ is the output of the $(l_{\textrm{e}}-1)$-th encoder layer at time $t$ and $\mathbf{z}^{l_{\textrm{e}}}\left(t-1\right)$ is the output of the encoder layer $l_{\textrm{e}}$ at time $t-1$. $\mathbf{C}^{l_{\textrm{e}}}\left(t-1\right)$ is the memory state of the $l_{\textrm{e}}$-th encoder layer at time $t-1$. Specifically, if $l_{\textrm{e}}=1$, the corresponding $\mathbf{z}^{0}$ represents the input sequence. In the decoding process, the output of the encoder $\mathbf{z}^{L_{\textrm{e}}}$ is fed as the input sequence $\mathbf{\hat{z}}^{0}$ to the decoder network. The output of the decoding function for $l_{\textrm{d}}\in\left\{ 1,2,\ldots,L_{\textrm{d}}\right\}$ at time $t$ is
\begin{equation}
\mathbf{\hat{z}}^{l_{\textrm{d}}}\left(t\right)=\varphi\left(\mathbf{w}^{l_{\textrm{d}}}\cdot\left[\hat{\mathbf{z}}^{l_{\textrm{d}}-1}\left(t\right),\hat{\mathbf{z}}^{l_{\textrm{d}}}\left(t-1\right),\mathbf{C}^{l_{\textrm{d}}}\left(t-1\right)\right]+\mathbf{b}^{l_{\textrm{d}}}\right)
%\mathbf{\hat{z}}^{l_{\textrm{d}}}=E\left(\hat{\mathbf{z}}^{l_{\textrm{d}}-1}\right)=\varphi\left(\mathbf{w}^{l_{\textrm{d}}}\cdot\hat{\mathbf{z}}^{l_{\textrm{d}}-1}+\mathbf{b}^{l_{\textrm{d}}}\right),
\end{equation}
where $\mathbf{\hat{z}}^{l_{\textrm{d}}}\left(t\right)$ and $\mathbf{\hat{z}}^{l_{\textrm{d}}}\left(t-1\right)$ is the output of the decoder of layer $l_{\textrm{d}}$ at time $t$ and $t-1$, respectively. $\hat{\mathbf{z}}^{l_{\textrm{d}}-1}\left(t\right)$ represents the output of the decoder layer $l_{\textrm{d}}-1$ at time $t$. Likewise, $\mathbf{w}^{l_{\textrm{d}}}$ and $\mathbf{b}^{l_{\textrm{d}}}$ represents the weight parameters and bias parameters of the $l_{\textrm{d}}$-th layer in decoder, respectively. $\mathbf{C}^{l_{\textrm{d}}}\left(t-1\right)$ is the memory state of the $l_{\textrm{d}}$-th decoder layer at time $t-1$. Moreover, $\mathbf{z}^{l_{\textrm{E}}}\left(t\right)$ the output of the encoder at time $t$ also represents the predicted vectors of the local/global popularities at time $t+1$, and will gradually approximate the true popularities along as the training.

\subsubsection{Distributed Training via FL}

The historical request data is denoted as ${\cal D} = \{ {{\cal D}^1},{{\cal D}^2}, \ldots ,{{\cal D}^I}\}$ where ${{\cal D}^i} = \{ {{ F}^i}(t)|t \in {\mathbb Z}_{0+} \}$ is the historical request data of UE-$i$ without labels. To address the privacy concern, the historical request data of each user is not exposed to others. During the offline training phase of URFL, UE-$i$ randomly extract $S$ samples from ${{\cal D}^i}$ using the extractor, denoted as ${{\cal T}^i}{\rm{ = }}\{ ({{{\bf{x}}_i}^s},{{{\bf{x}}_i}^s})
|{{{\bf{x}}_i}^s} = {{\bf{R}}^i}({t_s}),s = 1,2, \ldots ,S\}$, where ${t_s}$ is the random sample points at time slot $t$. Then, the training data is fed to the local AE in a mini-batch to train the network. The mini-batch average of the MSE loss function is adopted to yield a more stable convergence. That is, for any mini-batch set $\{ ({{{\bf{x}}_i}^{{s_w}}},{{{\bf{x}}_i}^{{s_w}}})|{{{\bf{x}}_i}^{{s_w}}} = {{\bf{R}}^i}({t_{{s_w}}}),{t_{{s_w}}} \in \{ {t_1},{t_2}, \ldots ,{t_S}\} ,w = 1,2, \ldots ,W\}$, we have
\begin{equation} \label{e17}
L({\Theta ^i}) = \frac{1}{W}\sum\nolimits_{w = 1}^W {{{\left| {{{\bf{R}}^i}({t_{{s_w}}}) - {{{\bf{\hat R}}}^i}({t_{{s_w}}})} \right|}^2}},
\end{equation}
where ${{{{\bf{\hat R}}}^i}({t_{{s_w}}})}$ is the output of the AE in UE-$i$. Note that, the above offline local training process is implemented in parallel in each UE.

\begin{algorithm}[!t]
\caption{URFL training for edge popularity prediction}
\label{alg_URFL}
\begin{algorithmic}[1]
 \STATE  \textbf{Initialization:} The extractor ${{\bf{R}}^i}(t)$ for each UE-$i$ and the global model parameters ${{\Theta ^{\rm{G}}}}$ are initialized randomly.
 \STATE \textbf{For} epoch $ t = 1, 2, \ldots, \Upsilon $ \textbf{do:}
\STATE \quad \quad The MEC server \textbf{do:}
\STATE \quad \quad \quad \quad \textbf{If} receive ${\Theta ^i}$ uploaded from users \textbf{then}
\STATE \quad \quad \quad \quad \quad Aggregate all the uploaded parameters ${\Theta ^{\rm{L}}}$ to \\
\quad \quad  \quad \quad \quad a global parameters set ${\Theta ^{\rm{AE}}}$ by \eqref{e18}.
\STATE \quad \quad \quad \quad \quad Update the global model parameters by \eqref{e19}.
\STATE \quad \quad \quad \quad \quad Broadcast ${\Theta ^{\rm{AE}}}$ to all users in its coverage.
\STATE \quad \quad \quad \quad \textbf{End If}
\STATE \quad \quad Each user $i \in {\cal I}$ \textbf{in parallel do:}

\STATE \quad \quad \quad \quad \textbf{If} receive ${\Theta ^{{\rm{AE}}}}$ broadcasted from the server \textbf{then}
\STATE \quad \quad \quad \quad \quad Update its local model parameters by\\
\quad \quad \quad \quad \quad \quad \quad \quad \quad \quad \quad ${\Theta ^i} = {\Theta ^{{\rm{AE}}}}$.
\STATE \quad \quad \quad \quad \textbf{End If}
\STATE \quad \quad \quad \quad Extract a mini-batch from ${{\cal D}^i}$ by extractor ${{\bf{R}}^i}(t)$.
\STATE \quad \quad \quad \quad Compute the MSE loss by \eqref{e17}, and update the\\
\quad \quad \quad \quad local model parameters ${\Theta ^i}$ using Adam.
\STATE \quad \quad \quad \quad \textbf{If} $t$ is an integer multiple of $T$ \textbf{then}
\STATE \quad \quad \quad \quad \quad Upload ${\Theta ^i}$ to the MEC server.
\STATE \quad \quad \quad \quad \textbf{End If}
\STATE \textbf{End For}
\end{algorithmic}
\end{algorithm}

Unlike the training process in the local UEs, the MEC server has no data to train its global prediction model under the constraint of the privacy-preserving mechanism stated in Section \ref{Prp_mec}. As such, we adopt the FL framework here to achieve the acquisition of local and global prediction models while preserving user privacy by aggregating parameters $\left\{{\Theta ^i}\right\}_{i = 1}^I$ instead of historical requests information. Concretely, by the end of every $T$ local self-training of the model ${f^{{\Theta ^i}}}( \cdot )$, UE-$i$ uploads its latest model parameters to the MEC server. Let ${\Theta ^{\rm{L}}} = \left\{{\Theta ^i}\right\}_{i = 1}^I = \left\{\Theta _{\rm{E}}^i,\Theta _{\rm{D}}^i\right\}_{i = 1}^I$ denote the stack of parameter sets uploaded by all the users. In the MEC server, the updated parameters set is aggregated from ${\Theta ^{\rm{L}}}$ as
\begin{equation} \label{e18}
{\Theta ^{{\rm{AE}}}}{\rm{ = }}\frac{1}{I}\sum\nolimits_{i = 1}^I {{\omega _i}{\Theta ^i}}  = \frac{1}{I}\left\{ {\sum\nolimits_{i = 1}^I {{\omega _i}\Theta _{\rm{E}}^i} ,\sum\nolimits_{i = 1}^I {{\omega _i}\Theta _{\rm{D}}^i} } \right\},
\end{equation}
where ${{\omega _i}}$ reflects the impact of each user's parameters in the aggregation. In this work, we assume that there is no priority among users. Hence, we reasonably set ${\omega _i} = 1,\forall i \in {\cal I}$ and, based on \eqref{e18}, we update the parameters of $f_{{\rm{FL}}}^{{\Theta ^{\rm{G}}}}( \cdot )$ by
\begin{equation} \label{e19}
{\Theta ^{\rm{G}}}{\rm{ = }}\frac{1}{I}\sum\nolimits_{i = 1}^I {{\omega _i}\Theta _{\rm{E}}^i}.
\end{equation}

Once the weight aggregation is complete, the new parameters ${\Theta ^{{\rm{AE}}}}$ will be broadcasted to all network users. For $\forall i \in {\cal I}$, UE-$i$ will immediately update its parameters by ${\Theta ^i}{\rm{ = }}{\Theta ^{{\rm{AE}}}}$. Then, UE-$i$ will train its local network again for another $T$ iterations. Upon completion, UE-$i$ will continue to upload the latest parameters to the MEC server. Then, a new loop is launched, and so forth \emph{ad infinitum}. The loop described above can also be named as a communication round in FL. The overall training process of the URFL algorithm is summarized in {\bf{Algorithm \ref{alg_URFL}}}. $\Upsilon$ is the total communication rounds between the local side and global side during the whole training. We also draw a schematic diagram of the parameters passing mechanism of URFL in Fig.~\ref{param_pass_URFL}. It should be noted that Fig.~\ref{param_pass_URFL} is given here to more clearly illustrate the parameter passing flow in the distributed FL framework during the training of the URFL algorithm, and the inputs and outputs of the neural networks are shown in Fig.~\ref{fig3}.

\begin{algorithm}[!t]
\caption{FedLWA for parameter aggregation}
\label{FedLWA_pa}
\begin{algorithmic}[1]
 \STATE  \textbf{Initialization:} Initialize ${\omega _i} = 1,\forall i \in {\cal I}$.
 \STATE Recieve parameters $\left\{ \Theta^{i}\right\} _{i=1}^{I}$ and losses $\left\{ L_{{\rm Avg}}\left(\Theta^{i}\right)\right\} _{i=1}^{I}$.
 \STATE Evaluate the normalized losses $\left\{ \gamma_{i}\right\} _{i=1}^{I}$ by
\[\gamma_{i}=L_{{\rm Avg}}\left(\Theta^{i}\right)\cdot\left(\sum\nolimits _{i=1}^{I}L_{{\rm Avg}}\left(\Theta^{i}\right)\right)^{-1}\]
\STATE Aggregate model parameters to $\Theta^{{\rm AE}}$ by \eqref{FedLWA}.
\STATE Obtain the updated parameters ${\Theta ^{\rm{G}}}$ by \eqref{FedLWA_G}.
\STATE Return parameters $\Theta^{{\rm AE}}$ and ${\Theta ^{\rm{G}}}$.
\end{algorithmic}
\end{algorithm}
\subsubsection{FedLWA for parameter aggregation}
Because of the non-i.i.d. user behaviors considered in this paper, the convergence of the LSTM-AE model on each local UE is inconsistent at the end of each communication round. Due to the fact that the convergence of one model can be reflected by its training loss, we therefore design a FedLWA parameter aggregation scheme on the basis of the proposed URFL algorithm to reduce the impacts of non-i.i.d. user behaviors. The URFL algorithm that applies the FedLWA-based parameter aggregation scheme is named FedLWA-based URFL algorithm. Concretely, the parameters passing during the training of the FedLWA-based URFL algorithm is basically identical to that of the URFL algorithm, except that each local user needs to additionally upload its average training loss ${L_{Avg}}\left( {{\Theta ^i}} \right)$ by the end of each $T$ local training. ${L_{Avg}}\left( {{\Theta ^i}} \right)$ can be expressed as:
\begin{equation} \label{lossavg}
{L_{\rm{Avg}}}\left( {{\Theta ^i}} \right) = \frac{1}{T}\sum\limits_{l = 1}^T {{L_l}\left( {{\Theta ^i}} \right)},
\end{equation}
where ${L_l}\left( {{\Theta ^i}} \right)$ calculated by \eqref{e17} is the loss value of user $i$ at the $l$-th training epoch in this communication round. We can observer from \eqref{e17} \eqref{lossavg} that ${L_{\rm{Avg}}}\left( {{\Theta ^i}} \right)$ contains no privacy information of user $i$. Thus, the upload of ${L_{\rm{Avg}}}\left( {{\Theta ^i}} \right)$ will not cause the leakage of user privacy.

At the MEC server side, weight for parameter aggregation in the FedLWA scheme is dependent on the convergence of each LSTM-AE model at current communication round, which is the normalized loss denoted as $\gamma_{i}=L_{{\rm Avg}}\left(\Theta^{i}\right)\cdot\left(\sum\nolimits _{i=1}^{I}L_{{\rm Avg}}\left(\Theta^{i}\right)\right)^{-1}$. Then, the FedLWA-based parameter aggregation can be written as:
\begin{equation} \label{FedLWA}
\Theta^{{\rm AE}}=\sum_{i=1}^{I}\omega_{i}\gamma_{i}\Theta^{i}.
\end{equation}
The parameters update of $f_{{\rm{FL}}}^{{\Theta ^{\rm{G}}}}( \cdot )$ can be rewritten as :
\begin{equation} \label{FedLWA_G}
\Theta^{{\rm G}}{\rm =}\sum\nolimits _{i=1}^{I}\omega_{i}\gamma_{i}\Theta_{{\rm E}}^{i}.
\end{equation}
The parameter aggregation process of the FedLWA scheme is summarized in {\bf{Algorithm \ref{FedLWA_pa}}}

\begin{algorithm}[!t]
\caption{URFL online prediction}
\label{online_URFL}
\begin{algorithmic}[1]
\STATE  \textbf{Initialization:} The extractor ${{\bf{R}}^i}(t)$ for each UE-$i$ and the ${{\bf{R}}^{\rm{G}}}(t)$ are initialized with zero array.
\STATE \textbf{For} time slot $ t = 1, 2, \ldots $ \textbf{do:}
\STATE \quad \quad The MEC server \textbf{do:}
\STATE \quad \quad \quad \quad Acquire ${{\bf{R}}^{\rm{G}}}(t)$ by receiving requests from users.
\STATE \quad \quad \quad \quad Acquire prediction ${{{\bf{\hat P}}}^{\rm{G}}}(t + 1)$ by feeding ${{\bf{R}}^{\rm{G}}}(t)$ \\
\quad \quad \quad \quad to the trained global prediction model.
\STATE \quad \quad \quad \quad Erase the private information ${{\bf{R}}^{\rm{G}}}(t)$.
\STATE \quad \quad Each UE-$i \in {\cal I}$ \textbf{in parallel do:}
\STATE \quad \quad \quad \quad Make a content request $F^{i}(t)\in\emptyset \cup\mathcal{F}$.
\STATE \quad \quad \quad \quad \textbf{If} $F^{i}\left(t\right)\notin\mathcal{C}_{i}(t)$ \textbf{then}
\STATE \quad \quad \quad \quad \quad Upload request $F^{i}\left(t\right)$ to the MEC server.
\STATE \quad \quad \quad \quad \textbf{End If}
\STATE \quad \quad \quad \quad Extract $\left\{ F^{i}(t-H+h)\right\} _{h=0}^{H}$ from ${{\cal D}^i}$ by ex-\\
\quad \quad \quad \quad tractor ${{\bf{R}}^i}(t)$.
\STATE \quad \quad \quad \quad Acquire prediction ${{{\bf{\hat P}}}^i}({\alpha ^i}(t + 1),t + 1)$ by feed-\\
\quad \quad \quad \quad ing ${{\bf{R}}^i}(t)$ to its trained local prediction model.
\STATE \textbf{End For}
\end{algorithmic}
\end{algorithm}

\subsubsection{Distributed Online Prediction}
During the online service phase, the edge popularity in the system at each time slot $t$ can be instantly predicted by evaluating \eqref{e11} and \eqref{e15}. Concretely, as shown in Fig. ~\ref{fig3}, the extractor ${{\bf{R}}^i}(t)$ of UE-$i$ firstly extracts the historical request information $\left[F^{i}(t-H),F^{i}(t-H+1),\cdots,F^{i}(t)\right]$ at time $t$. Then, $\left[F^{i}(t-H),F^{i}(t-H+1),\cdots,F^{i}(t)\right]$ will be fed into the LSTM-AE network of UE-$i$, and the output of the encoder ${{{\bf{\hat P}}}^i}({\alpha ^i}(t + 1),t + 1)$ is the popularity prediction of user $i$ at time $t+1$. At the MEC server side, the received request information from local users ${{\bf{R}}^{\rm{G}}}(t)$ will be fed into the trained global prediction model whose output ${{{\bf{\hat P}}}^{\rm{G}}}(t + 1)$ is the prediction of global popularity at time $t+1$. Moreover, to preserve users' privacy, the request information ${{\bf{R}}^{\rm{G}}}(t)$ will be erased as soon as it is fed to the global prediction model. The overall online prediction process of the URFL algorithm is summarized in {\bf{Algorithm \ref{online_URFL}}}.

\subsubsection{Time Complexity Analysis}
Finally, we investigate the time complexity of the proposed URFL algorithm from the perspective of local side and global side. Note that each UE holds local model with the same structure and executes the algorithm in parallel, thus the time complexity of the algorithm at the local side can be analysed from the local model on a single UE. Moreover, we can observe from Algorithm \ref{alg_URFL} and Algorithm \ref{online_URFL} that the whole structure of the LSTM-AE model participates in the training while only the encoder component of the LSTM-AE model participates in the online prediction. As such, the time complexity of the URFL algorithm at the local side during the training and prediction can be respectively expressed as $O\left(\sum_{l_{e}=1}^{L_{e}}HD_{l_{e}}^{2}+\sum_{l_{d}=1}^{L_{d}}HD_{l_{d}}^{2}\right)$ and $O\left(\sum_{l_{e}=1}^{L_{e}}HD_{l_{e}}^{2}\right)$, where $D_{l_{e}}$ and $D_{l_{d}}$ respectively denote the representation dimension of $l_{e}$-th encoder layer and $l_{d}$-th decoder layer in the LSTM-AE architecture. The time complexity of the URFL algorithm at the global side during the training comes from the parameter aggregation by \eqref{e19}, and thus can be denoted by $O\left(I\right)$. During the online prediction, the time complexity of the URFL algorithm at the global side mainly arises from the process of global popularity prediction and can be denoted by $O\left(\sum_{l_{e}=1}^{L_{e}}ID_{l_{e}}^{2}\right)$. Moreover, we can find from Algorithm \ref{FedLWA_pa} that the time complexity arises from the FedLWA parameter aggregation scheme is negligible compared with that of the URFL algorithm.
%Note that the resources of devices are sufficient during the offline training phase since devices are normally idle at this time. Therefore, we are more concerned with the computational complexity of the online prediction phase.

\section{NUMERICAL SIMULATIONS}
\label{sec4}

In this section, we showcase the superior performance of the proposed URFL algorithm in predicting the edge popularity. We run our numerical simulations on a workstation equipped with an Intel Xeon Gold 5118 CPU with 12 cores running at $2.30\;{\rm{GHz}}$ and $125\;{\rm{GB}}$ of RAM memory. The models and networks are trained and tested in the TensorFlow
environment. In the simulation, the window length of the extractor is $H = 10$. We assume that all the UEs have equal cache capacity denoted as $M_{i}=M_{j}, \forall i,j\in\mathcal{I}$.

In the trials, the parameter set ${{\cal G}_i}$ and the transition probability matrix $\mathbf{P}_{i}=\left\{P_{g_{l}g_{k}}^{i}\right\}_{g_{l},g_{k}=0}^{G_{i}}$ of each UE-$i$ are generated randomly, where $P_{g_{l}g_{k}}^{i}$ represents the transition probability from $\alpha_{g_{l}}^{i}$ to $\alpha_{g_{k}}^{i}$. In particularly, the entries of $\mathbf{P}_{i}$ can be arbitrary values, as it has no impact on the algorithm performance. As such, the statistical properties of the user behaviors are non-i.i.d. Under these parameters, users record their requests ${F^i}(t)$ over a period of time. Then, each UE-$i$ randomly extracts $S$ samples $({{{\bf{x}}_i}^s},{{{\bf{x}}_i}^s})$ from its own request record as the training dataset ${{\cal T}^i}$. In addition, we evaluate the performance of the proposed URFL algorithm on small groups of users, i.e., $I\in\{3,6,10\}$, which is also adopted in \cite{WangJ20,ZhangZ21}. Adam optimizer \cite{King14} is used to train the parameters $\left\{\Theta^{i}\right\}_{i=1}^{I}$ with an identical adaptive learning rate starting from $10^{-4}$. It should be emphasized that ${{\cal G}_i}$ and $\mathbf{P}_{i}$ are merely assigned to establish a similar-to-real simulation environment, neither of them are known to the MEC server and the UE-$i$ itself. For the architecture of the deep encoder $\Theta_{\textrm{E}}^{i}$ in the URFL, we use three hidden layers with $128$, $64$, $24$ LSTM neurons as well as a dropout rate of $0.35$ in each hidden layer to avoid overfitting. A mirror-symmetrical structure of the encoder is implemented in the decoder $\Theta_{\textrm{D}}^{i}$. Detailed simulation parameters are listed in Table \ref{simu_setup}.
\begin{table}[!t]
\centering
\caption{Simulation Parameters} \label{simu_setup}
\begin{tabular}{c|c}
\toprule[1pt]
    \textbf{Parameter} & \textbf{ Value} \\
    \hline\hline
     Total content number $N$ &  $12,18,24,32,38,44,50$ \\
    \hline
%     Observation window length $H $ & $2,4,6,8,10,12,14,16,18,20$ \\
%    \hline
%     User number $I$ & $3,6,10$ \\
%    \hline
     \tabincell{c}{Parameters Aggregation\\ coefficient of UE-$i$ ${{\omega _i}}$} & 1\\
    \hline
     Encoder hidden layer number $L_{\rm e}$ &  $3$ \\
    \hline
     Decoder hidden layer number $L_{\rm d}$ &  $3$ \\
     \hline
      \tabincell{c}{Number of local-training \\ epochs pre round $T$} & $8,16,32,64,128,256$  \\
     \hline
      \tabincell{c}{Samples $S$} & $100,1000,10000,100000$  \\
     \hline
      Learning rate for training & $10^{-4}$  \\
     \hline
      Optimizer for training & Adam \cite{King14} \\
     \hline
      Dropout rate & $0.35$  \\
     \hline
      \tabincell{c}{Content request arrival\\ rate of UE-$i$ $\lambda_{i}\left(t\right)$} & Randomly generated\\
    \hline
     \tabincell{c}{Local popularity distribution  \\parameter set of UE-$i$ ${{\cal G}_i}$} & Randomly generated \\
    \hline
     \tabincell{c}{Transition probability \\matrix of UE-$i$ $\mathbf{P}_{i}$} & Randomly generated \\
    \hline
\bottomrule[1pt]
\end{tabular}
\end{table}

\subsection{Baselines}

In the numerical simulations, we compare the prediction performance of the proposed URFL method with the baseline methods. All the reference methods are simulated in a $10$-user case. In addition, $2752$ epochs are run for all the learning methods.

\subsubsection{Singular Value Decomposition (SVD)}

A traditional and widely-adopted method in recommendation systems, i.e., SVD \cite{Zeyd16} is included in the comparison. The SVD method is centrally deployed on the MEC server without any privacy-preserving constraints. On the other hand, the privacy information of all the users can be accessed by the MEC server for training this baseline.

\subsubsection{Deep Recurrent AE Learning (DRAEL)}

We also consider an unsupervised learning method \cite{Hint06}, DRAEL, to evaluate the impacts of the FL modules in the proposed framework. For fairness, the AE architecture of the DRAEL is identical to that of the proposed URFL. Nevertheless, due to the removal of the FL framework in DRAEL, centralized training by feeding the private historical requests information of users is needed to train this method on the MEC server.

\subsubsection{Single Dense AE Federated Learning (SDAEFL) and Deep Dense AE Federated Learning (DDAEFL)}

We also consider other two distributed learning methods, SDAEFL and DDAEFL, to demonstrate the gains of the LSTM cells in the proposed URFL. More specifically, the FL framework is still used in these two methods to protect data privacy. However, the neural networks of the AEs in these two baseline methods are single dense neural networks and deep dense neural networks, respectively. Moreover, the encoder and decoder of the SDAEFL method are both single dense neural networks, and the number of hidden layers of AE in DDAEFL is equal to that of the proposed URFL.

\subsubsection{Self-train}
To evaluate the performance of the proposed URFL method at the local user side, we set a self-train method as another baseline method for comparison. This baseline method is deployed on the local UEs and has the same AE architecture as the proposed URFL method. Then, each UE trains its own local prediction model in parallel without any communications with other UEs or the MEC server.

\begin{table*}[!htbp]
\centering
\caption{Communication Cost Statistics} \label{tab_commcost}
\begin{threeparttable}
\begin{tabular}{@{}cccccc@{}}
\toprule[1pt]
\textbf{Type} & \textbf{Methods} & \textbf{\tabincell{c}{Total Data Traffic\\ (offline training)}} & \textbf{\tabincell{c}{Data Traffic per Time Slot \tnote{1}\\ (online prediction)}} & \textbf{\tabincell{c}{Prediction Error\\ (Global, RMSE)}} & \textbf{\tabincell{c}{Privacy \\ Preservation}} \\
\midrule \midrule
\multirow{2}{*}{\textbf{Centralized}}   & SVD & $267.03\; \rm{MB}$ & $40\; \rm{B}$ & $0.574$ & No \\
                                        & DRAEL & $267.03\; \rm{MB}$ & $40\; \rm{B}$ & $0.476$ & No \\
\midrule
\multirow{3}{*}{\textbf{Distributed}} & SDAEFL & $85.63\; \rm{MB}$ & $40\; \rm{B}$ & $0.588$ & Yes \\
                                        & DDAEFL & $211.73\; \rm{MB}$ & $40\; \rm{B}$ & $0.571$ & Yes \\
                                        & \textbf{URFL} (Proposed) & $853.74\; \rm{MB}$  & $40\; \rm{B}$ & 0.185 & Yes \\
\bottomrule[1pt]
\end{tabular}
 \begin{tablenotes}
        \footnotesize
        \item[1]  The data traffic pre time slot of each method is a theoretical value under the assumption that $\forall i\in\mathcal{I},C_{i}\left(t\right)=\emptyset,\lambda_{i}\left(t\right)=1$.
 \end{tablenotes}
\end{threeparttable}
\end{table*}

\begin{figure*}[!htpb]
    \centering
    \subfigure[]{\includegraphics[width=1\columnwidth,trim=3 2 40 40,clip]{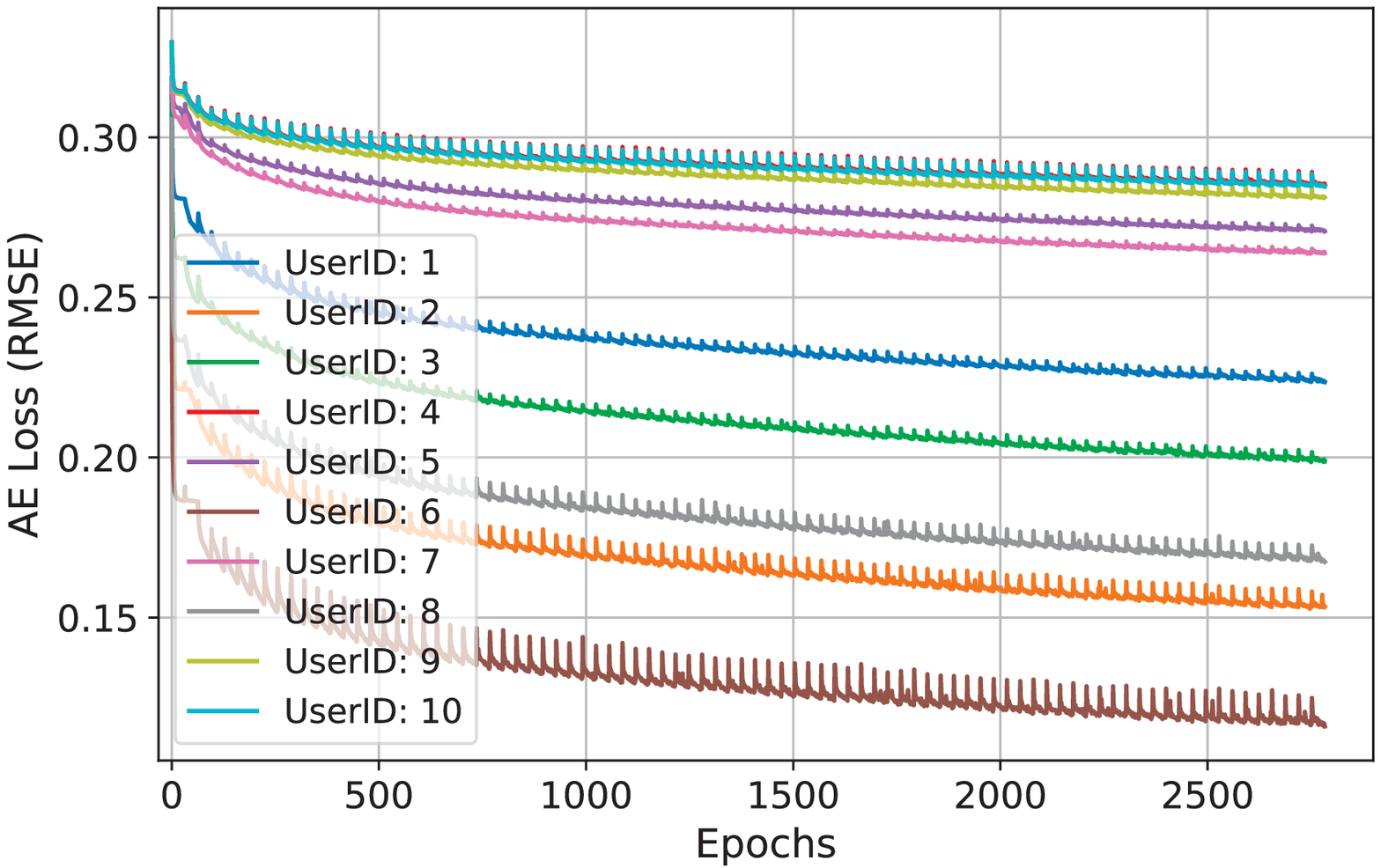}}
    \subfigure[]{\includegraphics[width=1\columnwidth,trim=3 2 40 40,clip]{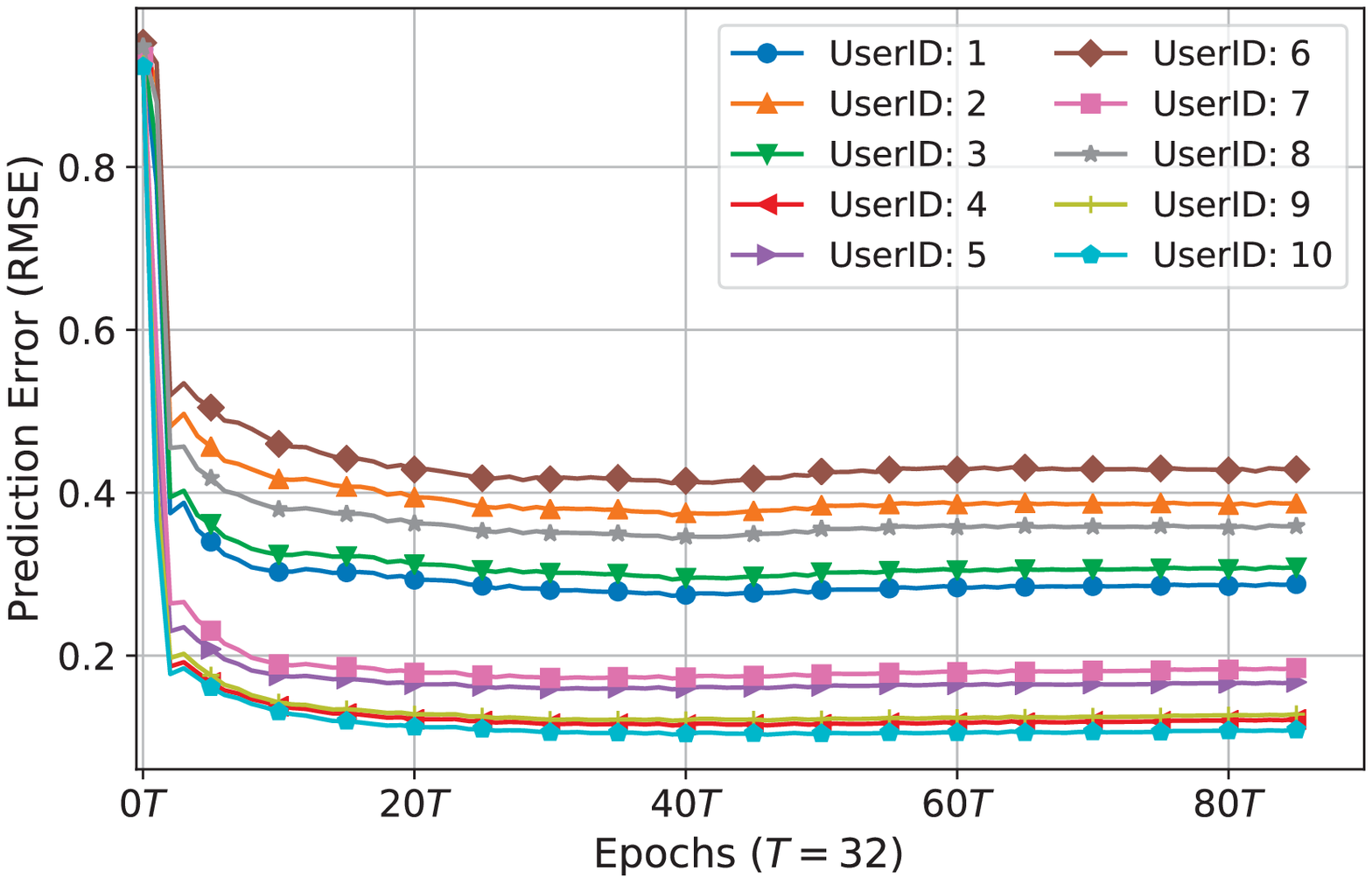}}
    \caption{Performance evaluations of the proposed URFL algorithm in the prediction of local popularity. (a) AE loss of each user. (b) Prediction error of each user.}
    \label{fig4}
\end{figure*}

\begin{figure}[!t]
\centering
\includegraphics[width=0.5\textwidth,trim=3 2 3 3,clip]{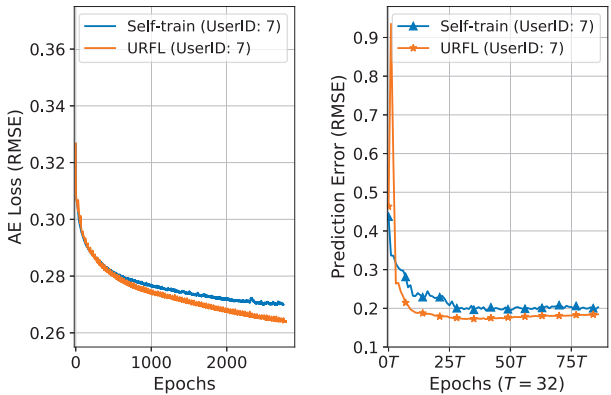}
\caption{Performance comparison of the proposed URFL method and self-train method on local popularity prediction, i.e., UserID $7$.}
\label{fig5}
\end{figure}
\subsection{Results and Discussions}

The prediction error measured by RMSE \cite{Sapu19} of the proposed algorithm in predicting the local popularity is shown in Fig. \ref{fig4}, where the prediction loss of each local model is sampled after every $T=32$ self-training iterations. It can be observed from Fig. \ref{fig4}(a) and (b) that, for each user, the prediction accuracy of the local popularity and the AE loss can both stably converge to a satisfactory level with the proposed approach. In particular, we identify many regular jitters on the RMSE curves as the learning epoch increases in Fig. \ref{fig4}(a). The reason is that all the local AEs are forced to aggregate their parameters based on FL after every $T$ rounds of self-training. As such, the RMSE loss of each local model instantaneously increases after the parametric aggregation of multiple UEs, and it then gradually decays to a lower level within the next $T$ self-training iterations. We also readily observe from Figs. \ref{fig4}(a) and (b) that, there are slight differences in the convergence losses of local popularity prediction for different users, which is reasonable since all the users are non-i.i.d. and both the set ${{\cal G}_i}$ and the probability $\mathbf{P}_{i}$ of each UE-$i$ are randomly generated. For the complicated ${{\cal G}_i}$ and $\mathbf{P}_{i}$, the challenge of prediction is tougher. In fact, this difference in prediction accuracy also indicates a non-i.i.d. open problem in FL \cite{Kair20}, which will be an important research focus in our future works.

We further take UserID 7 as an instance and examine the performance comparison of the proposed algorithm with the self-training method, which is commonly adopted in deep learning related works. In such a method, the agents train their own local models individually without any interactions with others. As shown in Fig.~\ref{fig5}, the proposed algorithm outperforms the self-training method from an individual user perspective in terms of AE loss and prediction loss. This result suggests that proper interactions with other participants can improve the prediction accuracy, which is congenial with common sense.
\begin{figure}[!t]
\centering
\includegraphics[width=0.46\textwidth]{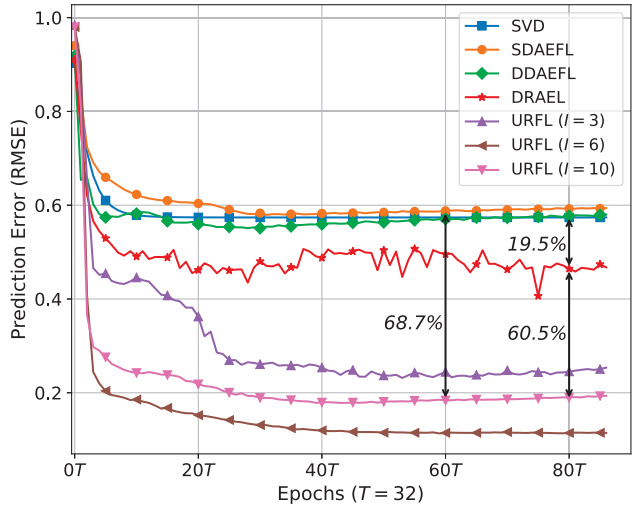}
\caption{Performance comparison of the baselines and proposed URFL method in the prediction of global popularity.}
\label{fig6}
\end{figure}

We take a step forward and study the prediction performance from the perspective of global popularity. Fig.~\ref{fig6} implies that the proposed method is superior to all the other baseline methods in terms of the prediction accuracy. In the $10$-UE case, the proposed method yields an RMSE of around $68.7\%$ lower than those of SVD, SDAEFL, and DDAEFL, and $60.5\%$ lower than that of DRAEL. The remarkable gain suggests that the proposed method can significantly improve the prediction accuracy while grantee the data privacy of users since it only aggregates the model parameters of each user rather than the private raw data. We argue that the aggregation process of the global prediction model pushes the MEC server to deeply and accurately learn the underlying features from all the UEs towards its coverage. In contrast, the baseline methods not only need to be supplied by large amounts of private data but also are incompetent to exact features from a mass of historical requests kneaded together in time and space. In addition, we observe from Fig.~\ref{fig6} that the performance of SDAEFL is inferior to SVD and DDAEFL, which implies that the single dense neural network cannot effectively predict popularity. Furthermore, the $19.5\%$ gain of the DRAEL to the DDAEFL and the $60.5\%$ gain of the proposal to the DRAEL confirm that the recurrent mode realized by LSTM and the parameters aggregation realized by FL can both contribute to the reduction of prediction error considerably. We also infer from the comparison between the proposed algorithm and DRAEL in Fig.~\ref{fig6} that, the prediction variance can be significantly reduced using the proposal. It is interesting to note, increasing the number of UEs does not continuously improve the prediction accuracy of the proposed algorithm. As can be seen from Fig.~\ref{fig6}, the best RMSE performance is achieved when the number of UEs $I=6$ rather than $I=10$ or $I=3$. This is because aggregation of insufficient local models can be unrepresentative of the global characteristics, while an overly large sample size will result in information redundancy.

\begin{figure}[!t]
\centering
\includegraphics[width=0.445\textwidth]{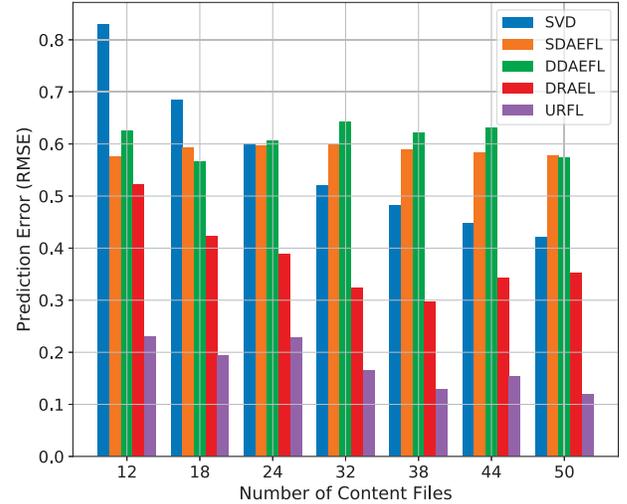}
\caption{Performance comparison of different number of contents files on the global prediction error.}
\label{fig_diffnum_contents}
\end{figure}

\begin{figure*}[!htpb]
    \centering
    \subfigure[]{\includegraphics[width=1\columnwidth,trim=3 2 40 40,clip]{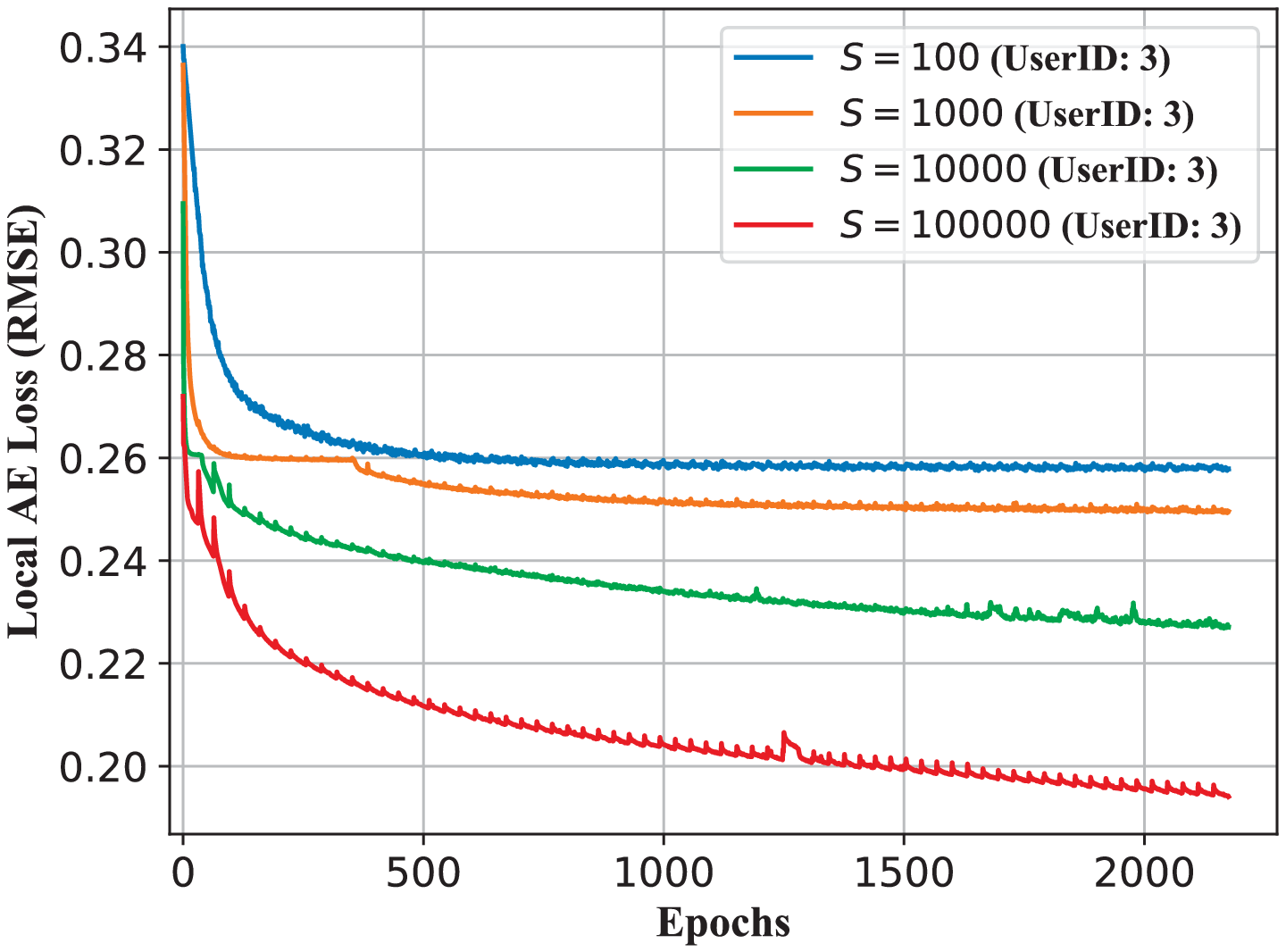}}
    \subfigure[]{\includegraphics[width=1\columnwidth,trim=3 2 40 40,clip]{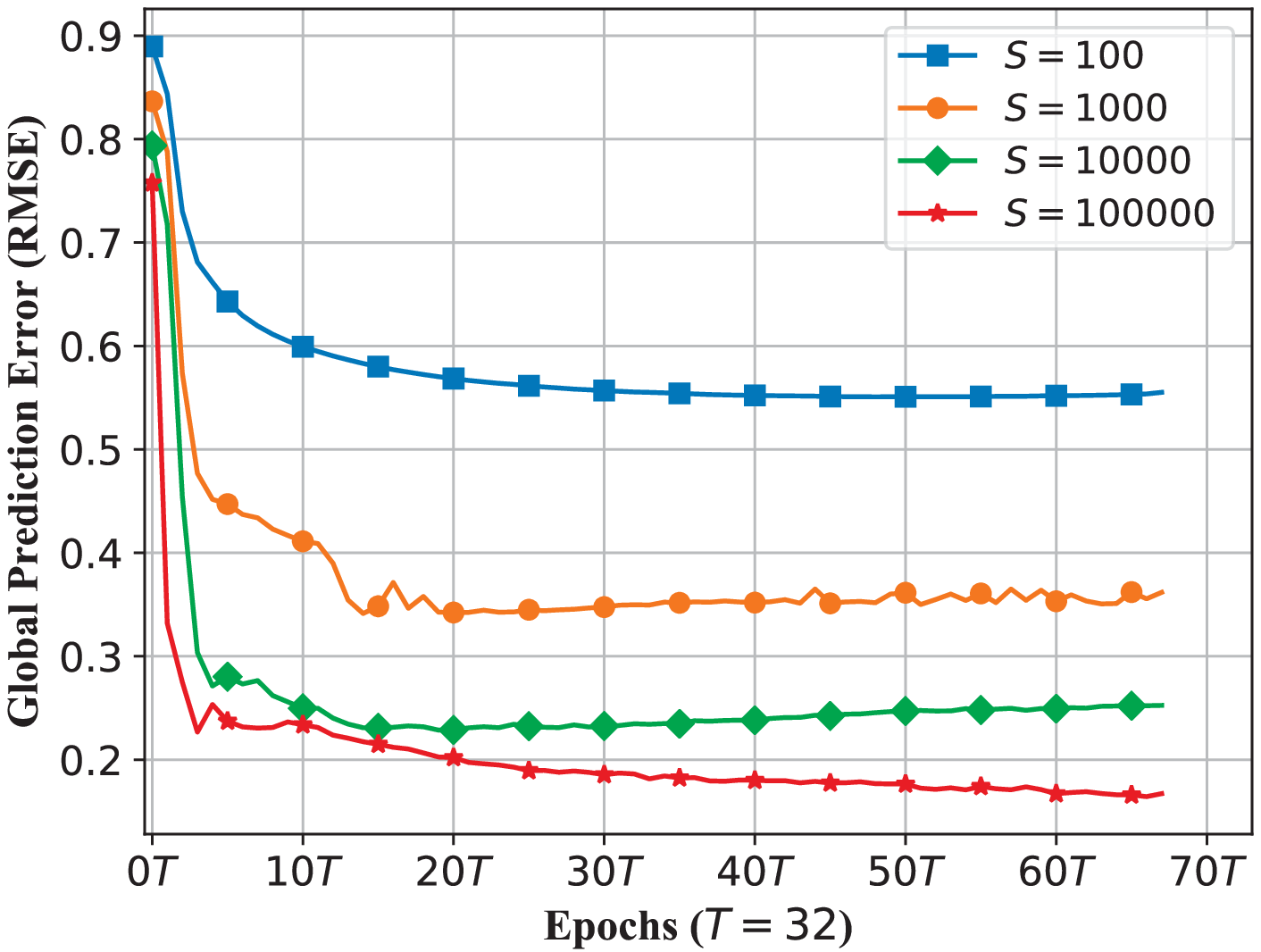}}
    \caption{Performance evaluations of the proposed URFL algorithm on different number of samples. ($I=3,N=24,H=10$) (a) AE loss of each local user. (b) Prediction error of the global popularity.}
    \label{diffsamples}
\end{figure*}
\begin{figure*}[!htpb]
    \centering
    \subfigure[]{\includegraphics[width=0.66\columnwidth]{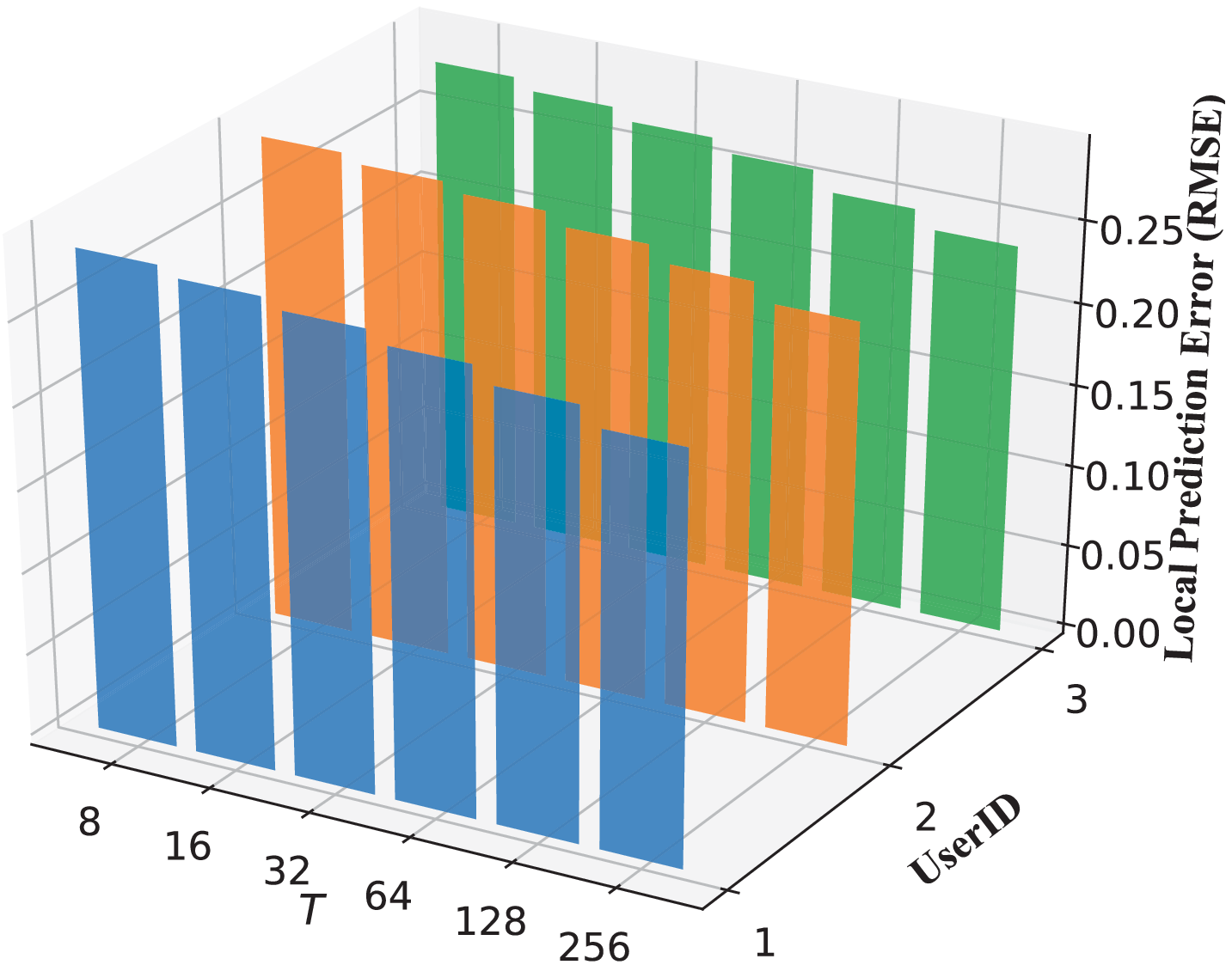}}
    \subfigure[]{\includegraphics[width=0.66\columnwidth]{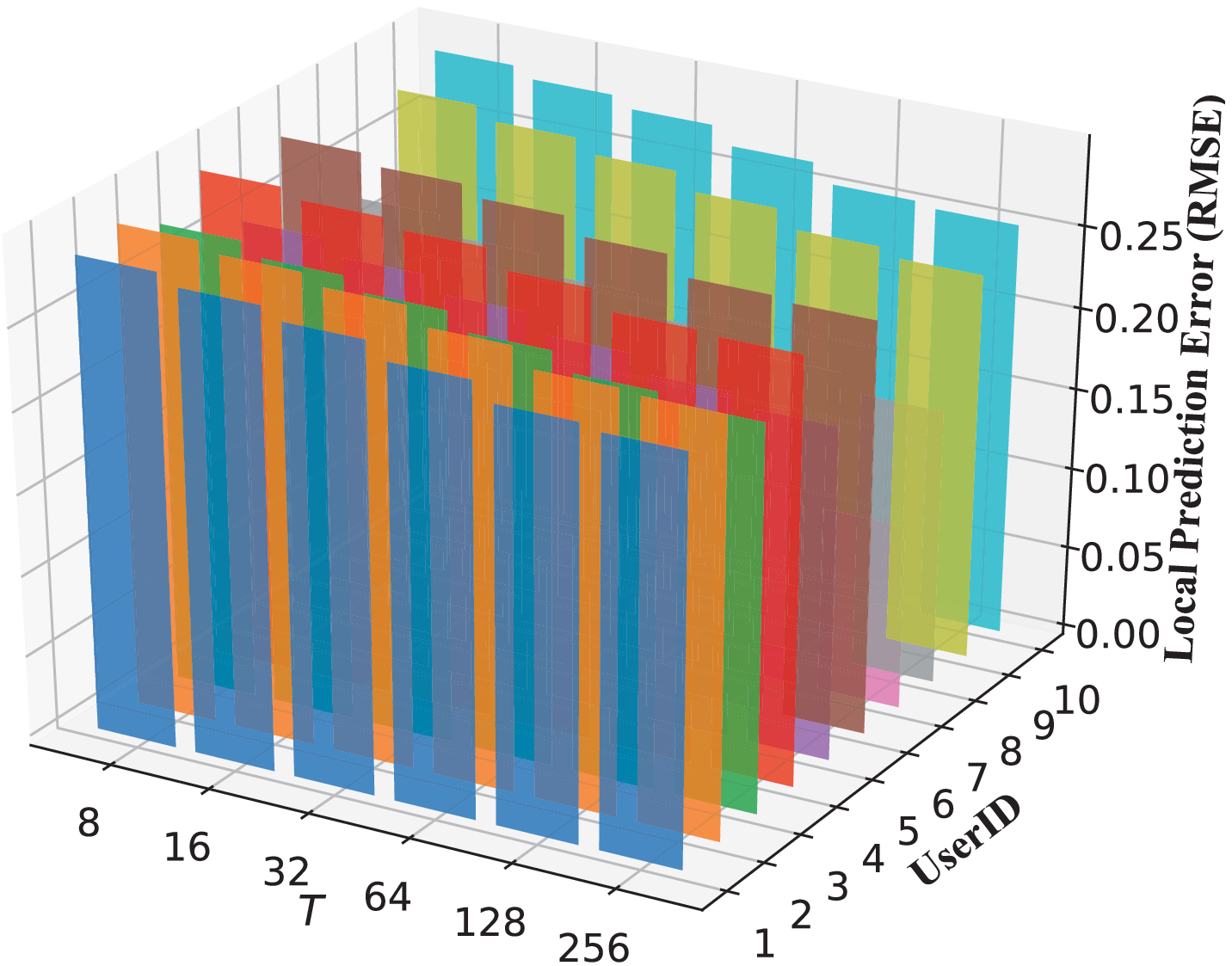}}
    \subfigure[]{\includegraphics[width=0.66\columnwidth,trim=3 2 40 40,clip]{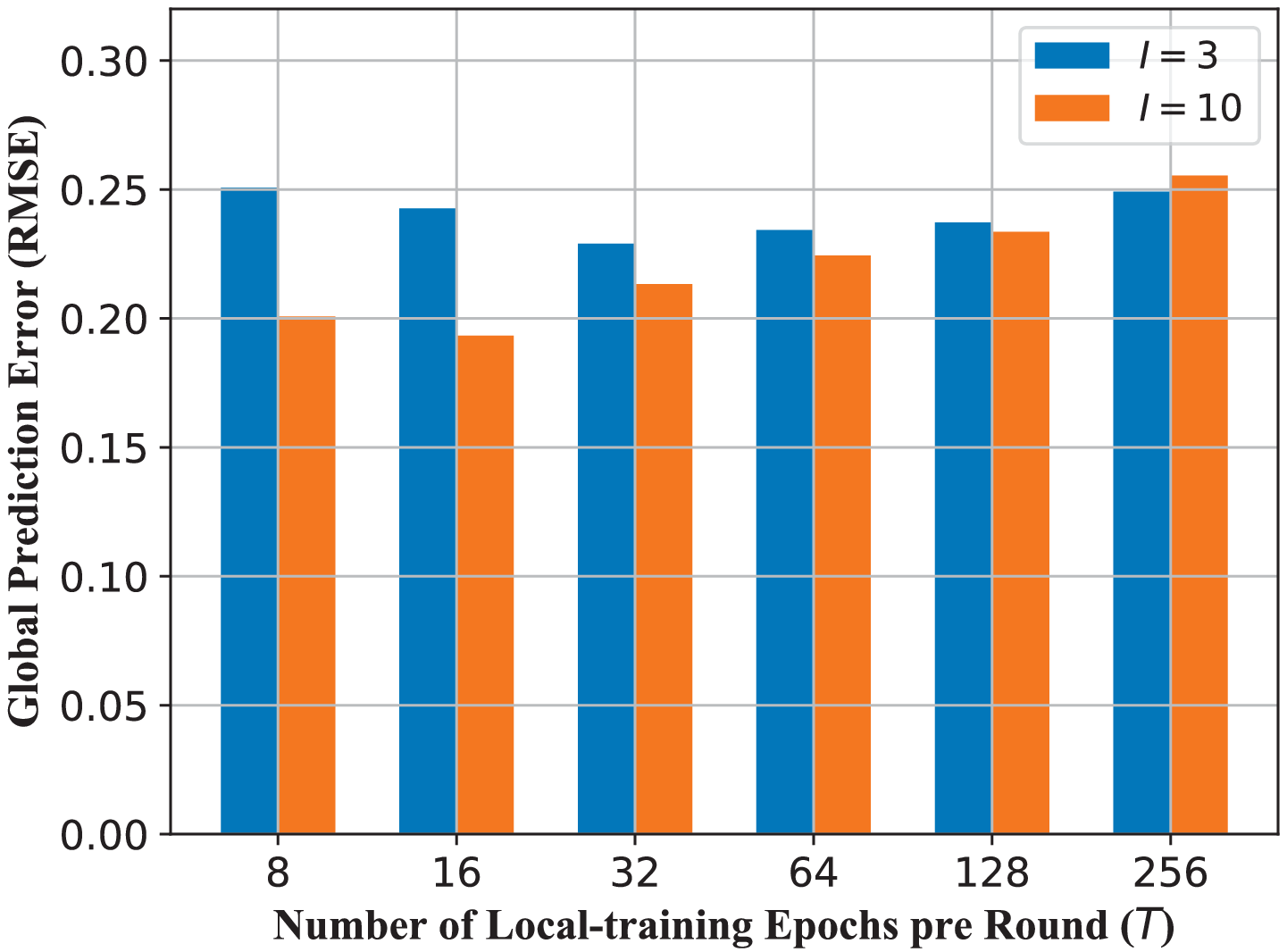}}
    \caption{Performance evaluations of the proposed URFL algorithm on different T. ($N=24,H=10$) (a) Prediction error of each local user. ($I=3$) (b) Prediction error of each local user. ($I=10$) (c) Prediction error of the global popularity.}
    \label{diffT}
\end{figure*}

The performance comparisons between the proposed method and all the baseline algorithms versus the number of total content files $N$ are presented in Fig. \ref{fig_diffnum_contents}. It can be seen from Fig. \ref{fig_diffnum_contents} that the proposed URFL algorithm significantly outperforms all the baselines regardless of the value of $N$. The statistics of communication cost represented by the data traffic in the MEC network are listed in Table \ref{tab_commcost}, where $I=10,N=24,T=32$. Though we can observe from Table \ref{tab_commcost} that the proposed URFL algorithm generates more data traffic for its offline training than the other baseline methods, the offline training phase is often implemented when the UE is idle, i.e., dormant status, charging status, etc. Therefore, it is acceptable for the proposed method to reduce the error of popularity prediction and preserve the privacy of users by sacrificing the tolerable increase of data traffic in the idle status of the MEC network. Moreover, under the assumption that $\forall i\in\mathcal{I},C_{i}\left(t\right)=\emptyset,\lambda_{i}\left(t\right)=1$, the theoretical data traffic of all the considered methods per time slot are equal, but the proposed method can achieve the lowest prediction error. When the assumption is invalid in the real environment, the data traffic of the centralized methods still remains $40\; \rm{B}$. By contrast, the data traffic of the distributed methods will be less than $40\; \rm{B}$, which actually depends on the $\lambda_{i}\left(t\right)$, prediction errors of local/global popularities, and the cache hit rates of each cache entities.

\begin{figure}[!t]
\centering
\includegraphics[width=0.455\textwidth]{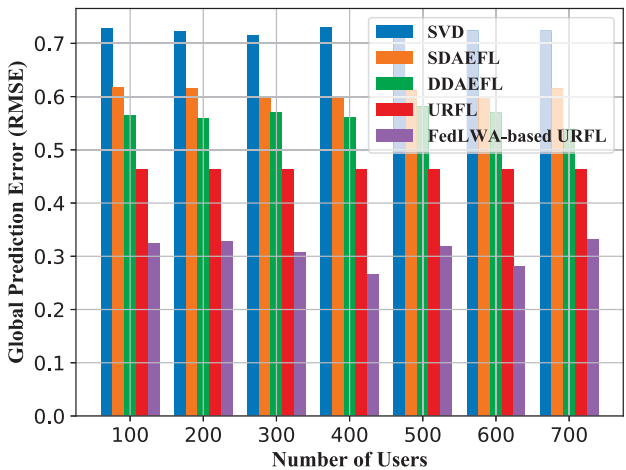}
\caption{Performance comparison on large gropus of users. ( $N=24,H=10,T=16$ )}
\label{fig_diffnum_UEs}
\end{figure}

\begin{figure*}[!htpb]
    \centering
    \subfigure[]{\includegraphics[width=1\columnwidth]{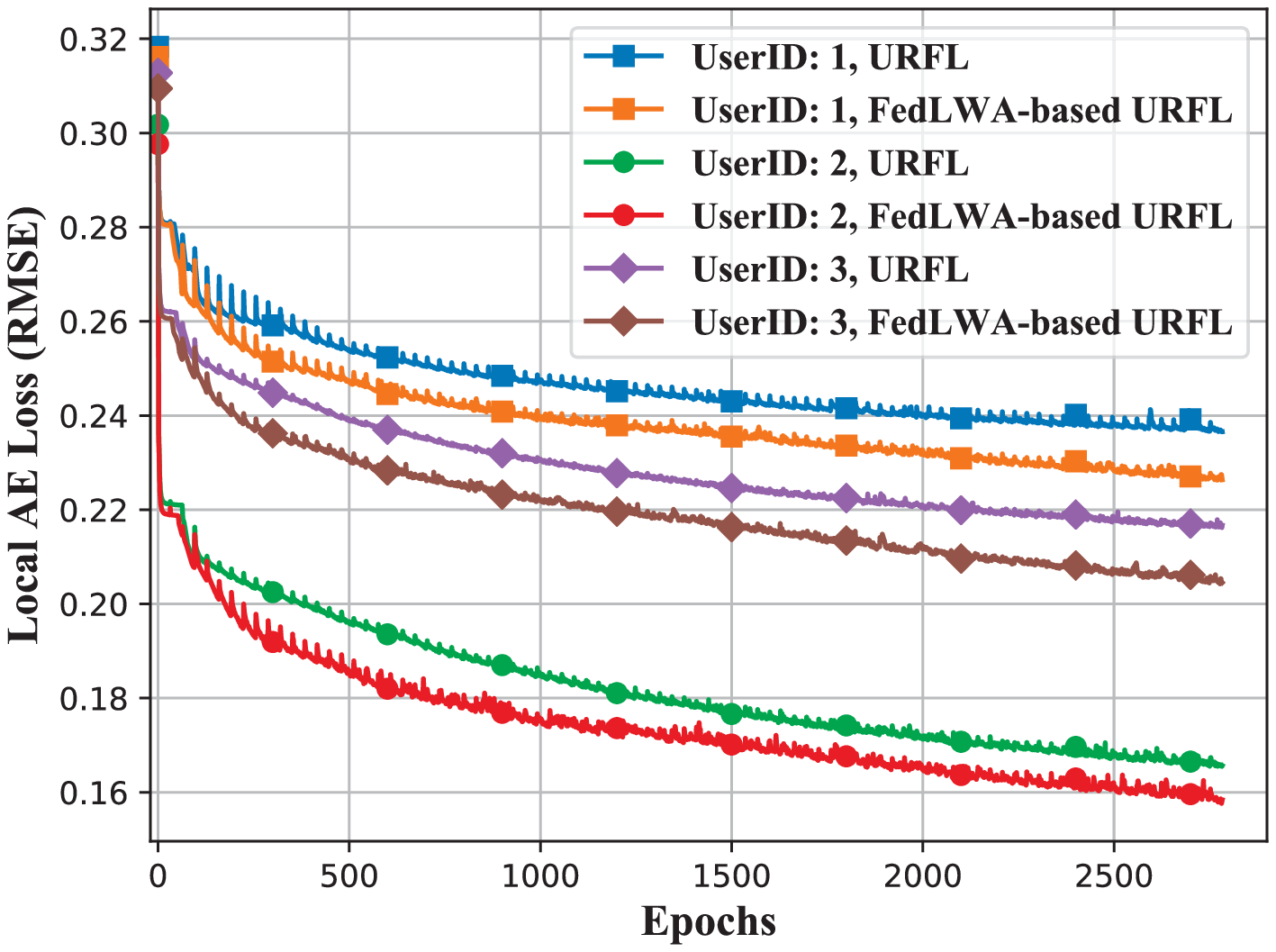}}
    \subfigure[]{\includegraphics[width=1\columnwidth]{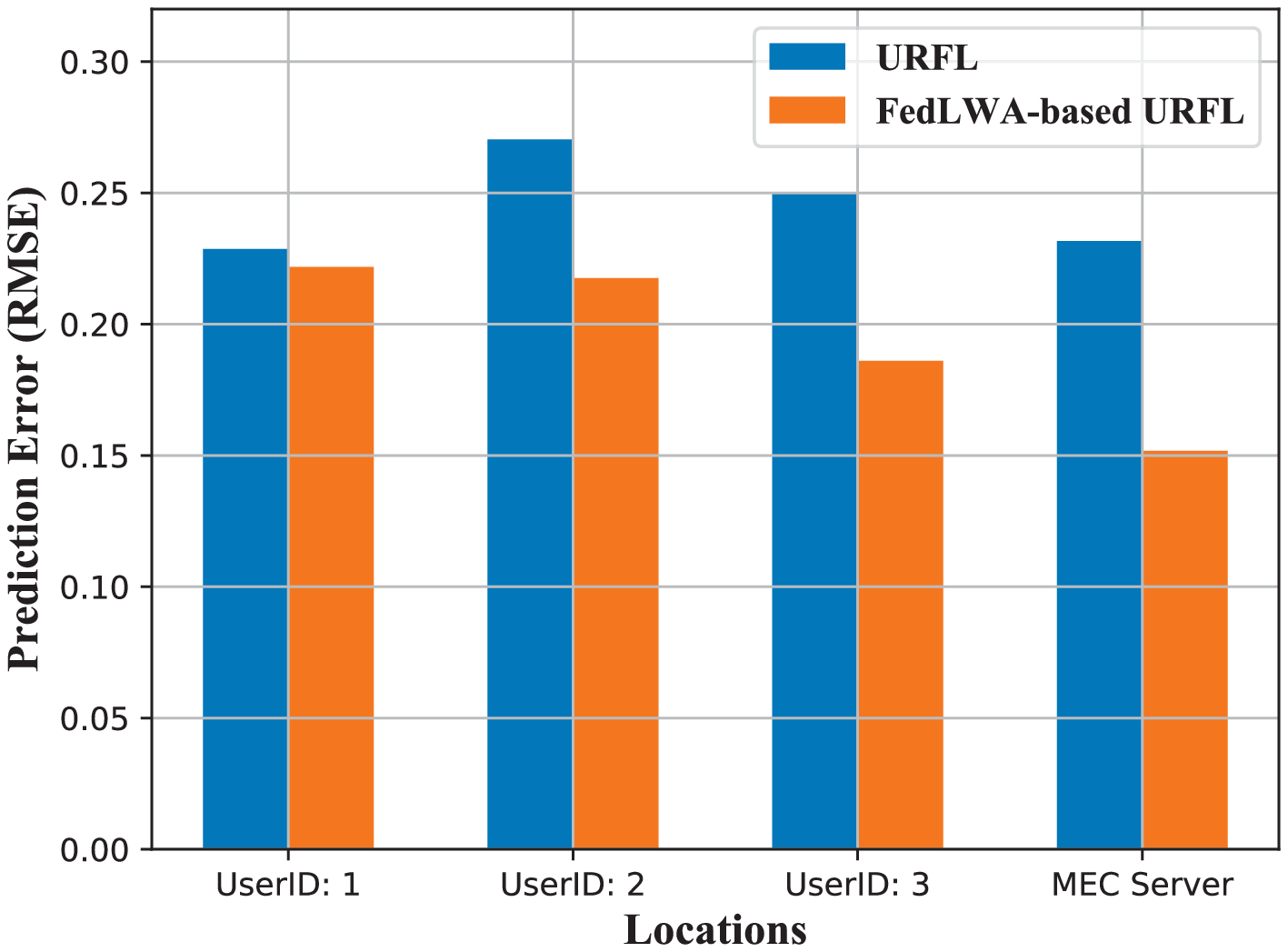}}
    \caption{Performance comparison of the proposed URFL algorithm and the proposed FedLWA-based URFL algorithm. ($N=24,H=10,I=3$) (a) AE loss of each user. (b) Prediction errors of local and global popularities.}
    \label{FedLWA_URFL_perf}
\end{figure*}
\begin{figure*}[!hbpt]
    \centering
    \subfigure[]{\includegraphics[width=0.67\columnwidth]{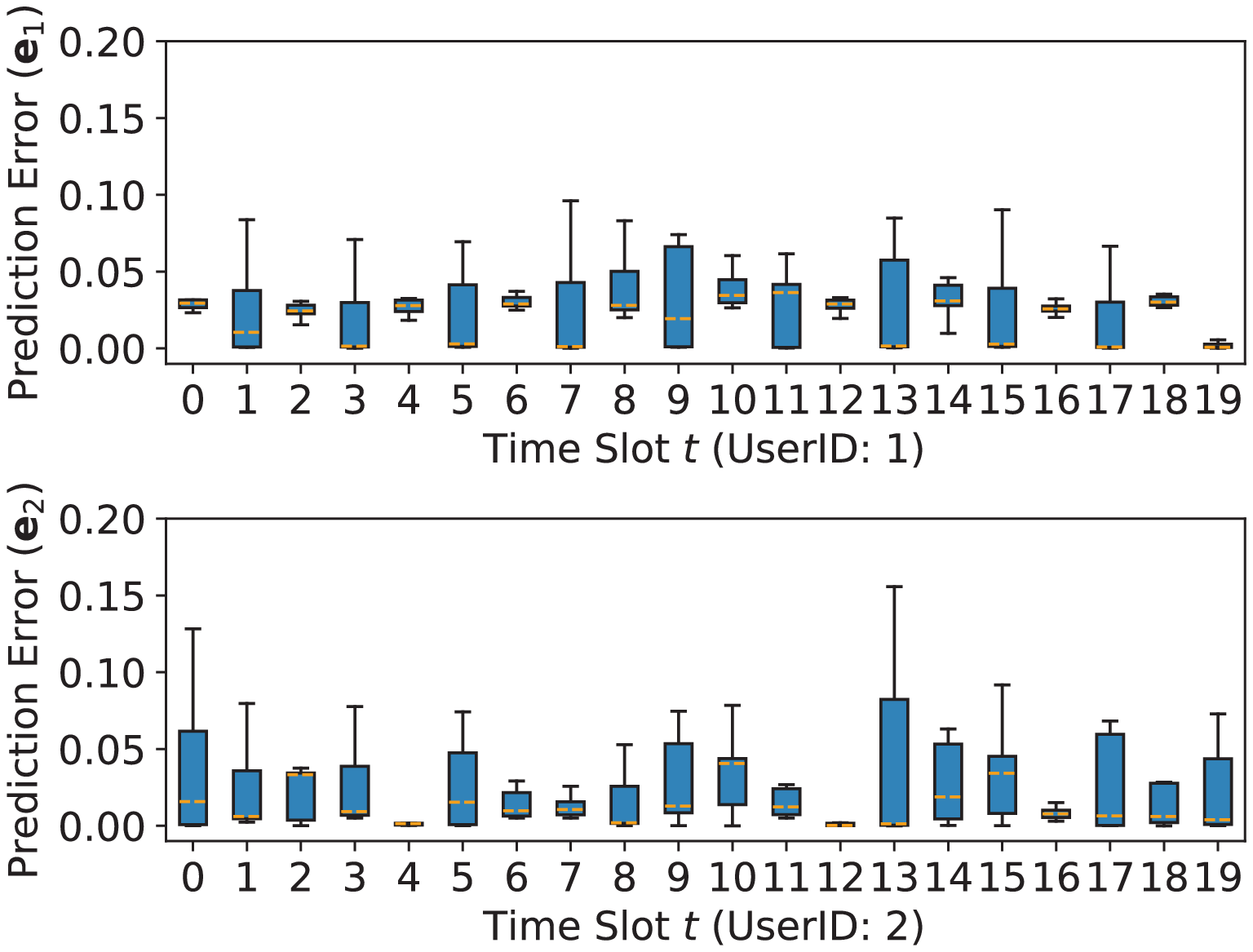}}%
    \hfil
    \subfigure[]{\includegraphics[width=0.67\columnwidth]{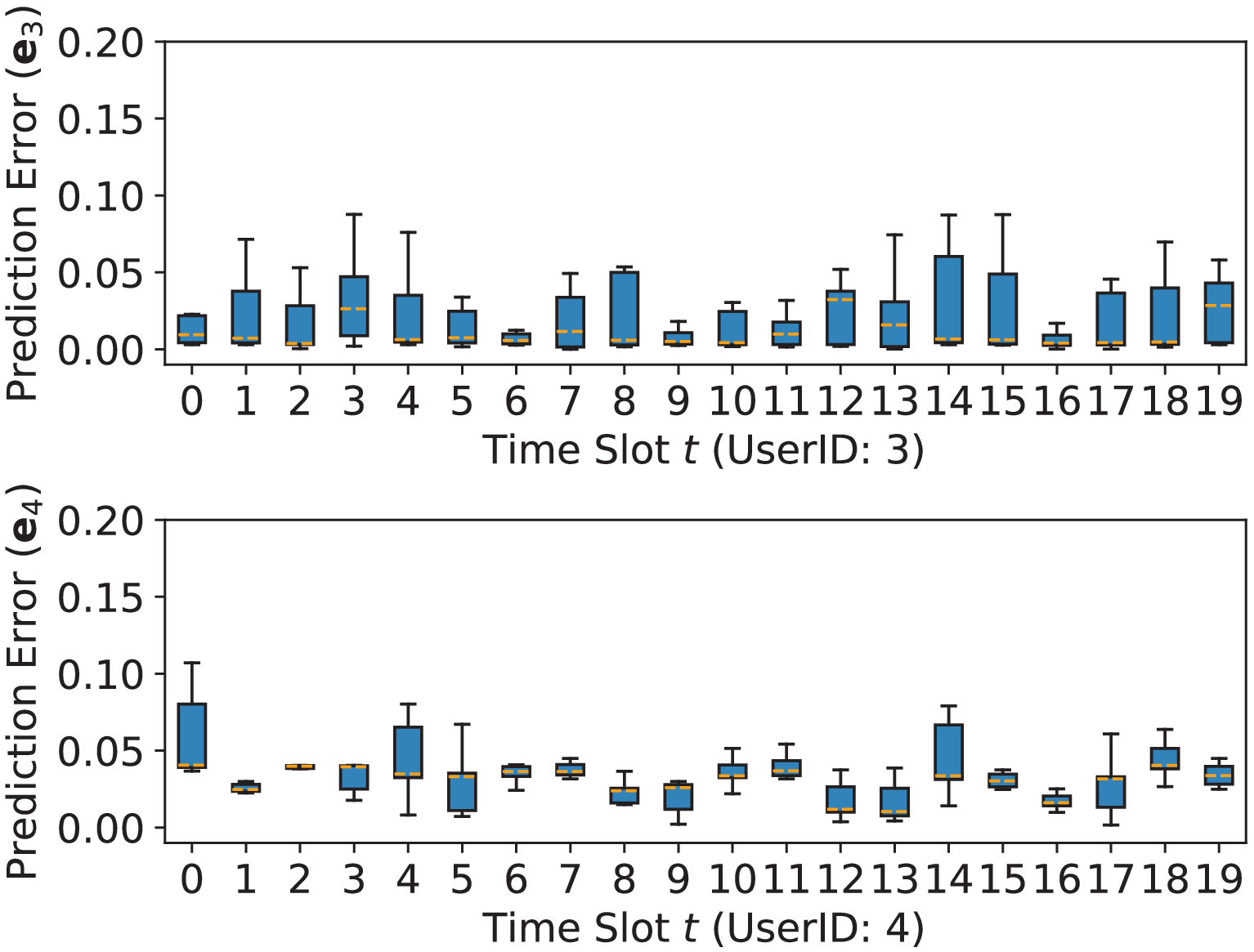}}%
    \hfil
    \subfigure[]{\includegraphics[width=0.67\columnwidth]{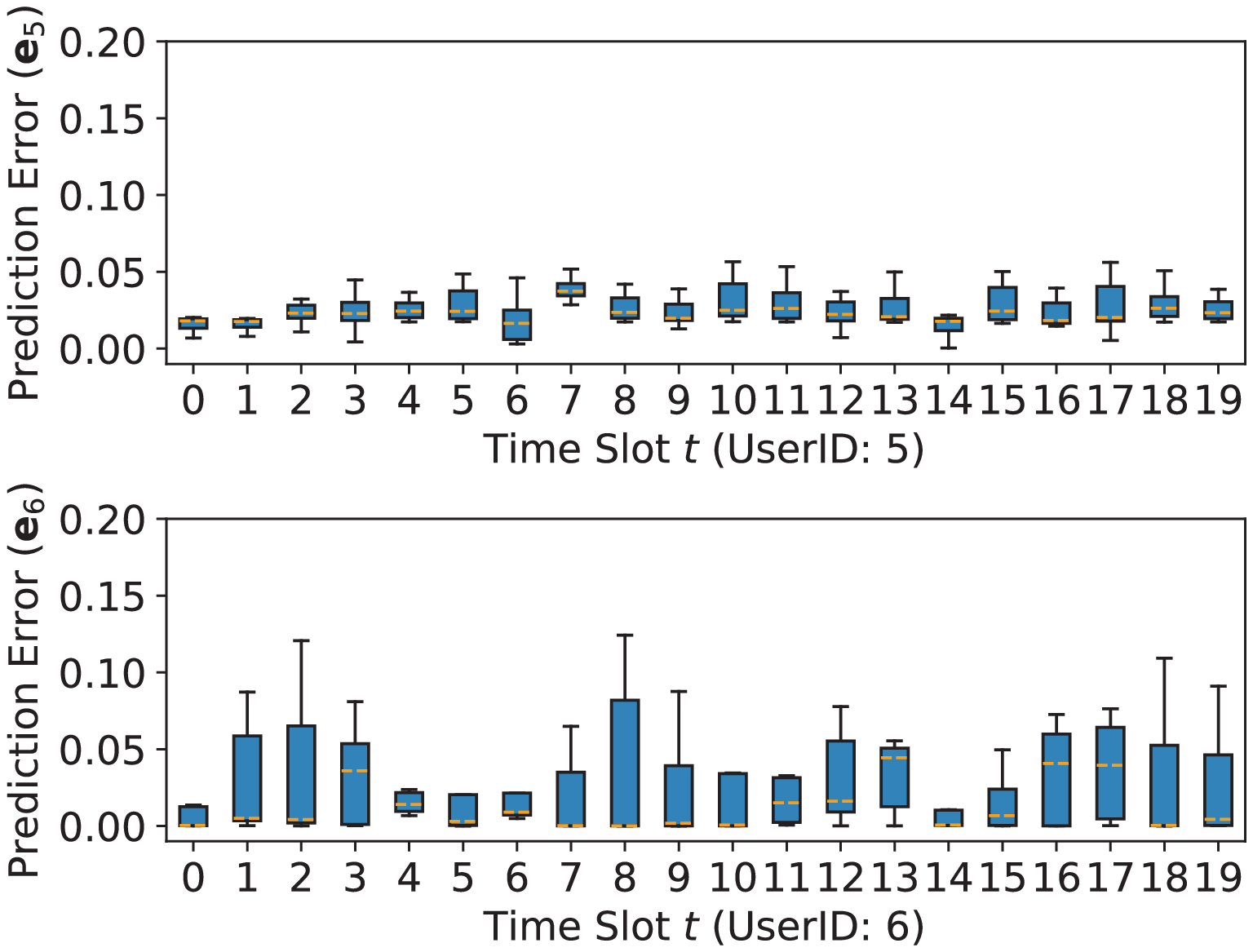}}\\
    \hfil
    \subfigure[]{\includegraphics[width=0.67\columnwidth]{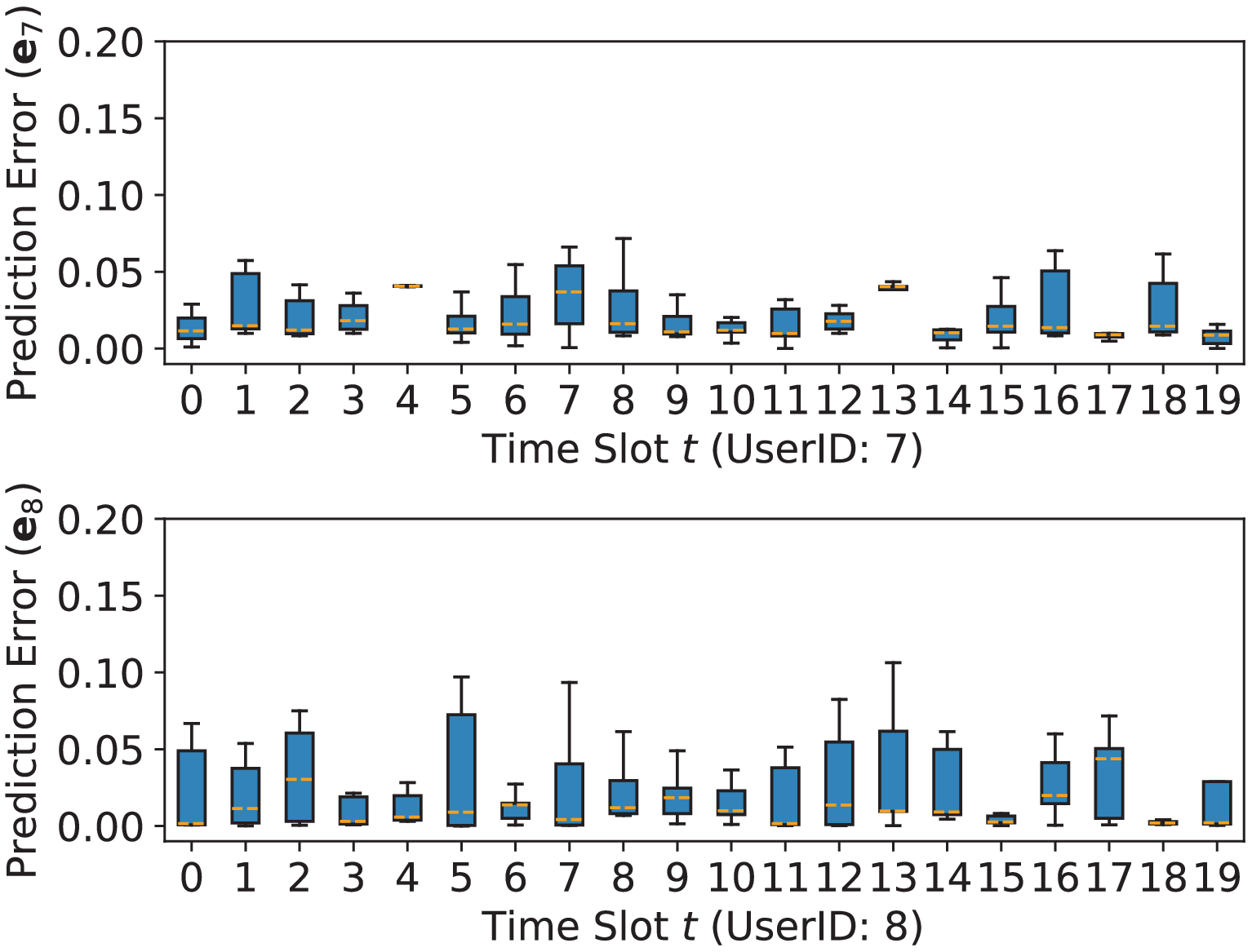}}%
    \hfil
    \subfigure[]{\includegraphics[width=0.67\columnwidth]{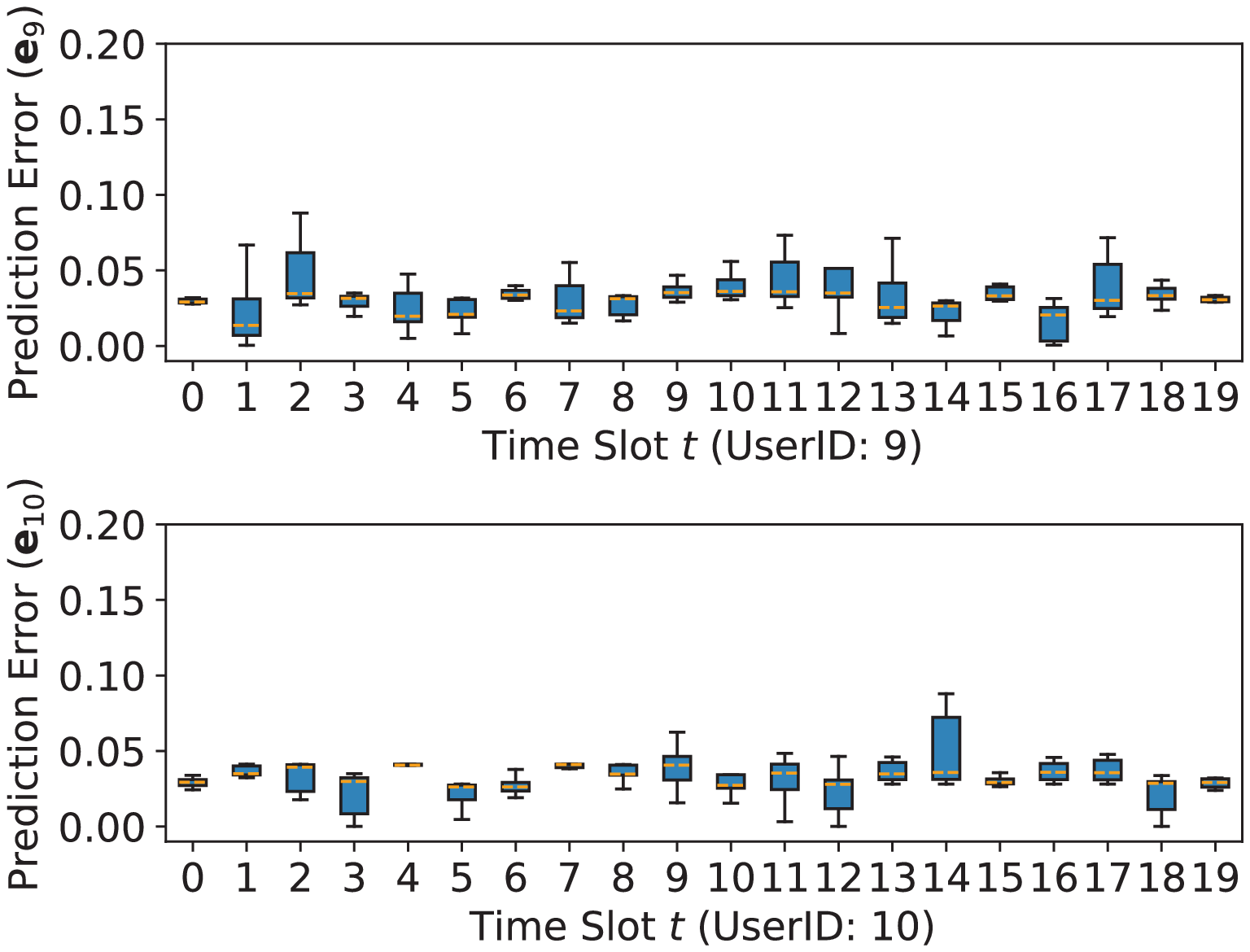}}%
    \hfil
    \subfigure[]{\includegraphics[width=0.68\columnwidth]{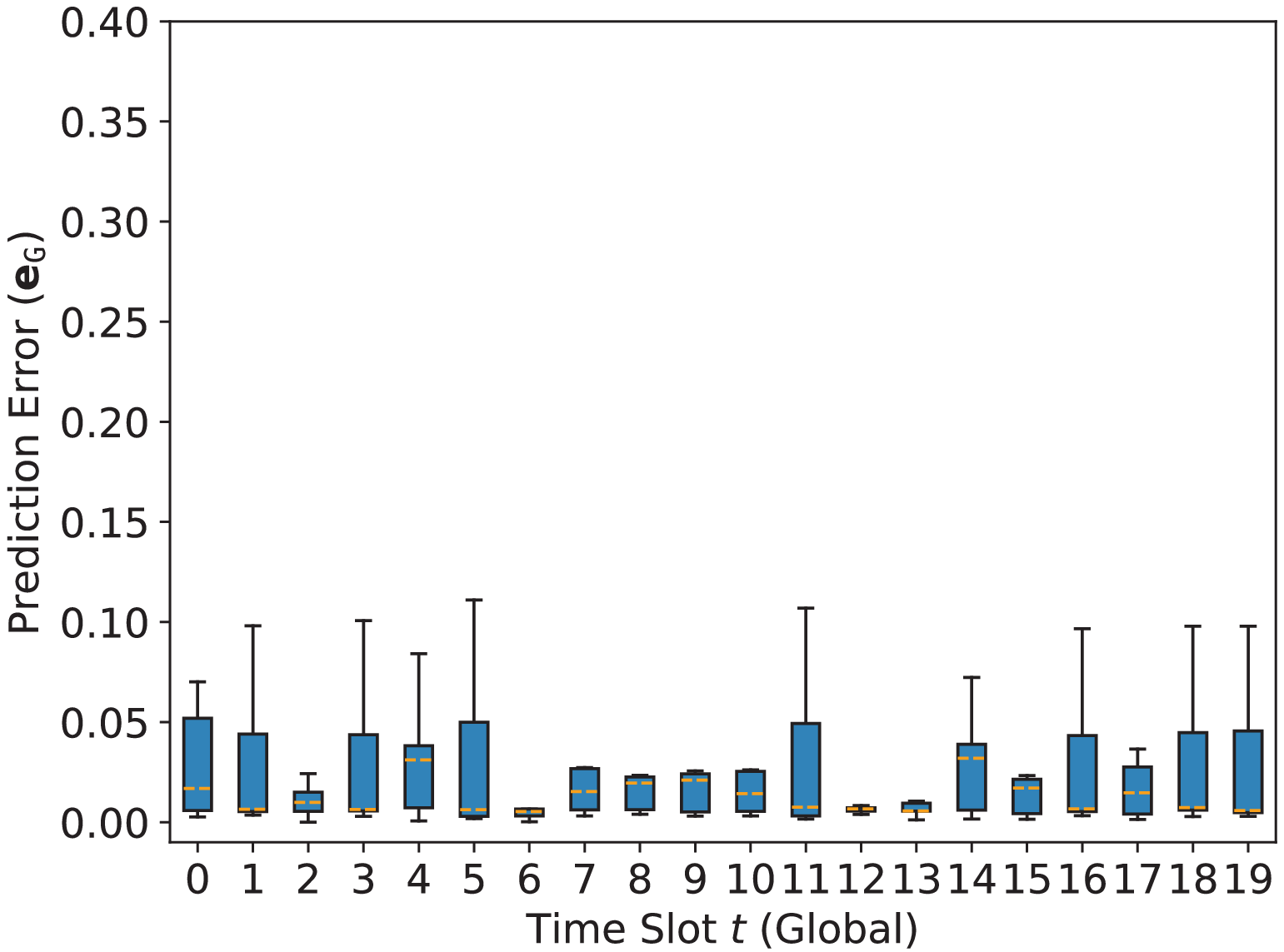}}
    \caption{ Performance test of the proposed URFL algorithm in the real-time prediction of local popularities and global popularity. (a) The real-time prediction error on UserID 1 and 2. (b) The real-time prediction error on UserID 3 and 4. (c) The real-time prediction error on UserID 5 and 6. (d) The real-time prediction error on UserID 7 and 8. (e) The real-time prediction error on UserID 9 and 10. (b) The real-time prediction error of the global popularity.}
    \label{fig7}
\end{figure*}

In Fig. \ref{diffsamples}, we provide the performance evaluations of the proposed URFL algorithm on different number of samples. We can observe from Fig. \ref{diffsamples}(a) and (b) that the proposed method can achieve superior performance in both local and global popularity predictions when the sample size is more sufficient. The reason is that the prediction model generally extracts more complete features from a dataset with sufficient samples, thus achieves the lower prediction error. With the sample size increasing to a certain extent, the further performance gains will not be produced due to the feature saturation. Moreover, we also evaluate the performance of the proposed URFL algorithm on different $T$, as shown in Fig. \ref{diffT}. Fig. \ref{diffT}(a) and (b) imply that the prediction error of the local popularity decreases with increasing $T$. This is because the local prediction errors are valued under the same communication rounds. As such, the local prediction model will be trained with more epochs as $T$ increases, and thus converges to a better performance level. Straightforward, if we want to achieve the the same local prediction performance under different $T$, we should continue increasing the number of communication rounds in the case of small $T$, but which also means higher communication costs. Fig. \ref{diffT} (c) illustrates that the prediction error of the global popularity will first decrease and then increase as $T$ increases. This performance trend is due to the fact that large local-training epochs pre round will bring a large bias between each local model, while few $T$ will lead to an under-optimization of each local model on their local dataset pre round.

In Fig. \ref{FedLWA_URFL_perf}, we provide the performance evaluation of the proposed FedLWA parameter aggregation scheme. It can be observered from Fig.\ref{FedLWA_URFL_perf}(a) that, for each user, the training loss of the LSTM-AE model can converge to a lower level with the proposed FedLWA-based URFL algorithm, which demonstrates that the FedLWA parameter aggregation scheme can bring additional performance gains to the training. The reason is that the proposed FedLWA scheme can adaptively weigh the contribution of each local parameter to the global parameter on the basis of the convergence of local models at the end of each communication round, so as to obtain better aggregated parameters. Consequently, as shown in Fig. \ref{FedLWA_URFL_perf}(b), the FedLWA-based URFL approach is superior to the URFL approach in terms of the prediction errors at both local user and MEC server sides. Hence, the non-i.i.d. problems in the investigated scenario could be alleviated by applying the proposed FedLWA parameter aggregation scheme.

The performance comparisons between the proposed methods and the baseline methods on large groups of users, i.e. $I=\left\{ 100,200,\cdots,700\right\} $, are presented in Fig. \ref{fig_diffnum_UEs}. Note that random sampling aggregation is a common approach to FL for large-scale client scenarios \cite{ZhangZ21}. Herein, we randomly select $6$ local clients for each parameter aggregation in the training phase. It can be observed from Fig. \ref{fig_diffnum_UEs} that, in large-scale user scenarios, the proposed URFL algorithm still outperform the baselines regardless of the value of $I$. Moreover, Fig. \ref{fig_diffnum_UEs} showcases the superiority of the proposed FedLWA-based URFL algorithm, which demonstrates that the proposed FedLWA parameter aggregation scheme is also applicable to large-scale user scenarios and can bring additional performance gains regardless of the number of users.

In closing, we test the online prediction performance of the proposed URFL algorithm in the local UEs and the MEC server in $20$ consecutive time slots. In the simulations, each UE and the MEC server are deployed with their own prediction models which have entered the convergent state using the proposed URFL algorithm. During the service delivery, the trained prediction models predict the future local and global popularities in real-time to assist the equipments to effectively update caches. In each time slot for arbitrary $i$, we compute the respective absolute errors between the predicted local popularity and the true local popularity of the request probability associated with each content file, denoted as $\mathbf{e}_{i}\left(t\right)=\left\{\left|\hat{P}_{n}^{i}\left(\alpha^{i}\left(t\right),t\right)-P_{n}^{i}\left(\alpha^{i}\left(t\right),t\right)\right|\right\}_{n=1}^{N}$. The real-time prediction errors of each UE are visualized in Figs. \ref{fig7}(a) to (e). Likewise, we evaluate the real-time prediction errors of the global popularity by $\mathbf{e}_{\textrm{G}}\left(t\right)=\left\{\left|\hat{P}_{n}^{i}\left(\alpha^{i}\left(t\right),t\right)-P_{n}^{i}\left(\alpha^{i}\left(t\right),t\right)\right|\right\}_{n=1}^{N}$ and show the result in Fig. \ref{fig7}(f). We readily observe from Fig. \ref{fig7} that, for both local and global popularities, the real-time prediction error for each content file is largely lower than $0.1$ and, in most of the cases, it is under $0.05$. This result shows that the convergent URFL algorithm can significantly reduce the prediction error of the local/global popularities during the online service, and this reconfirms the effectiveness of the proposed algorithm.

\begin{table*}[!htbp]
\scriptsize
\centering
\caption{User parameters for Simulation validation of Theorem \ref{theorem1}} \label{statis_set}
\begin{tabular}{@{}cccccccccccccccccccc@{}}
\toprule[1pt]
\multirow{2}{*}{\textbf{Distributions}} & \multicolumn{18}{c}{\textbf{User Parameters}}& \\ \cmidrule(lr){2-19}
 & \multicolumn{3}{c}{\textbf{UserID 1}} & \multicolumn{3}{c}{\textbf{UserID 2}} & \multicolumn{3}{c}{\textbf{UserID 3}} & \multicolumn{3}{c}{\textbf{UserID 4}} & \multicolumn{3}{c}{\textbf{UserID 5}} & \multicolumn{3}{c}{\textbf{UserID 6}}   \\ \midrule[0.65pt]\midrule[0.65pt]
\multirow{2}{*}{${\rm{Zipf}}\left(\alpha^{i}\right)$} & $\lambda_{1}$ & \multicolumn{2}{c}{$\alpha^{1}$} & $\lambda_{2}$ & \multicolumn{2}{c}{$\alpha^{2}$} & $\lambda_{3}$ & \multicolumn{2}{c}{$\alpha^{3}$} & $\lambda_{4}$ & \multicolumn{2}{c}{$\alpha^{4}$} & $\lambda_{5}$ & \multicolumn{2}{c}{$\alpha^{5}$} & $\lambda_{6}$ & \multicolumn{2}{c}{$\alpha^{6}$}  \\ \cmidrule(lr){2-19}
 & 0.74 & \multicolumn{2}{c}{0.08} & 0.91 & \multicolumn{2}{c}{2.14} & 0.58 & \multicolumn{2}{c}{1.56} & 0.76 & \multicolumn{2}{c}{1.02} & 0.74 & \multicolumn{2}{c}{0.11} & 0.63 & \multicolumn{2}{c}{0.15} &  \\ \midrule[0.65pt]
\multirow{2}{*}{$\rm{Poisson}\left(l^{i}\right)$} & $\lambda_{1}$ & \multicolumn{2}{c}{$l^{1}$} & $\lambda_{2}$ & \multicolumn{2}{c}{$l^{2}$} & $\lambda_{3}$ & \multicolumn{2}{c}{$l^{3}$} & $\lambda_{4}$ & \multicolumn{2}{c}{$l^{4}$} & $\lambda_{5}$ & \multicolumn{2}{c}{$l^{5}$} & $\lambda_{6}$ & \multicolumn{2}{c}{$l^{6}$}  \\ \cmidrule(lr){2-19}
 & 0.51 & \multicolumn{2}{c}{8} & 0.60 & \multicolumn{2}{c}{27} & 0.68 & \multicolumn{2}{c}{24} & 0.96 & \multicolumn{2}{c}{29} & 0.98 & \multicolumn{2}{c}{13} & 0.79 & \multicolumn{2}{c}{11} &  \\ \midrule[0.65pt]
\multirow{2}{*}{$\rm{nBernoulli}\left(p^{i}\right)$} & $\lambda_{1}$ & \multicolumn{2}{c}{$p^{1}$} & $\lambda_{2}$ & \multicolumn{2}{c}{$p^{2}$} & $\lambda_{3}$ & \multicolumn{2}{c}{$p^{3}$} & $\lambda_{4}$ & \multicolumn{2}{c}{$p^{4}$} & $\lambda_{5}$ & \multicolumn{2}{c}{$p^{5}$} & $\lambda_{6}$ & \multicolumn{2}{c}{$p^{6}$}  \\ \cmidrule(lr){2-19}
 & 0.94 & \multicolumn{2}{c}{0.44} & 0.91 & \multicolumn{2}{c}{0.11} & 0.91 & \multicolumn{2}{c}{0.50} & 0.68 & \multicolumn{2}{c}{0.70} & 0.76 & \multicolumn{2}{c}{0.52} & 0.70 & \multicolumn{2}{c}{0.51} &  \\ \midrule[0.65pt]
\multirow{2}{*}{$\rm{Gaussian}\left(\mu^{i},\sigma^{i}\right)$} & $\lambda_{1}$ & $\mu^{1}$ & $\sigma^{1}$ & $\lambda_{2}$ & $\mu^{2}$ & $\sigma^{2}$ & $\lambda_{3}$ & $\mu^{3}$ & $\sigma^{3}$ & $\lambda_{4}$ & $\mu^{4}$ & $\sigma^{4}$ & $\lambda_{5}$ & $\mu^{5}$ & $\sigma^{5}$ & $\lambda_{6}$ & $\mu^{6}$ & $\sigma^{6}$  \\ \cmidrule(lr){2-19}
 & 0.88 & 6 & 2.30 & 0.82 & 31 & 3.63 & 0.97 & 17 & 2.45 & 0.87 & 28 & 2.96 & 0.68 & 15 & 3.27 & 0.94 & 9 & 5.37 &  \\ \bottomrule[1pt]
\end{tabular}
\end{table*}

\begin{figure*}[!hbpt]
    \centering
    \subfigure[]{\includegraphics[width=0.5\columnwidth]{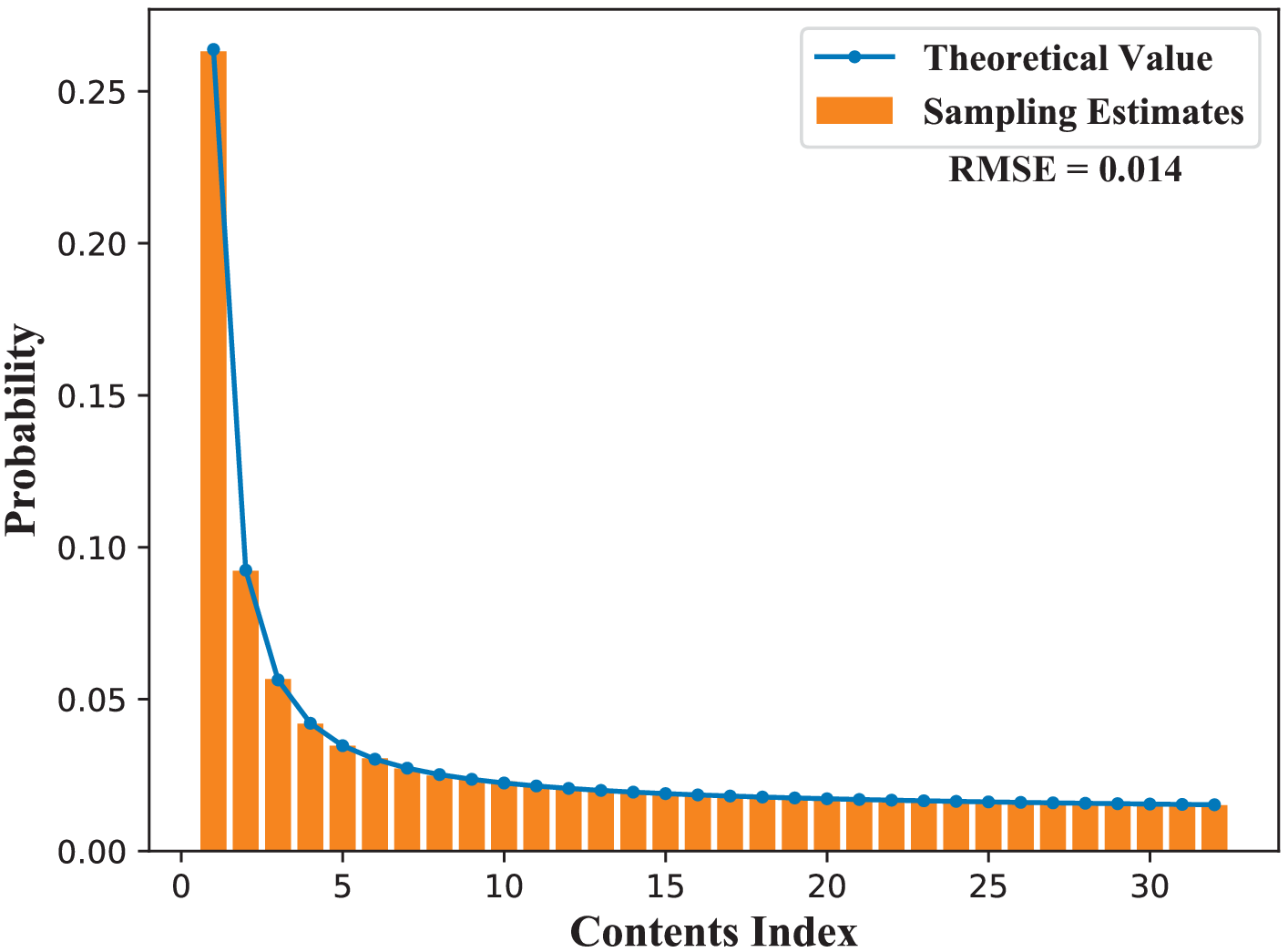}}
    \hfil
    \subfigure[]{\includegraphics[width=0.5\columnwidth]{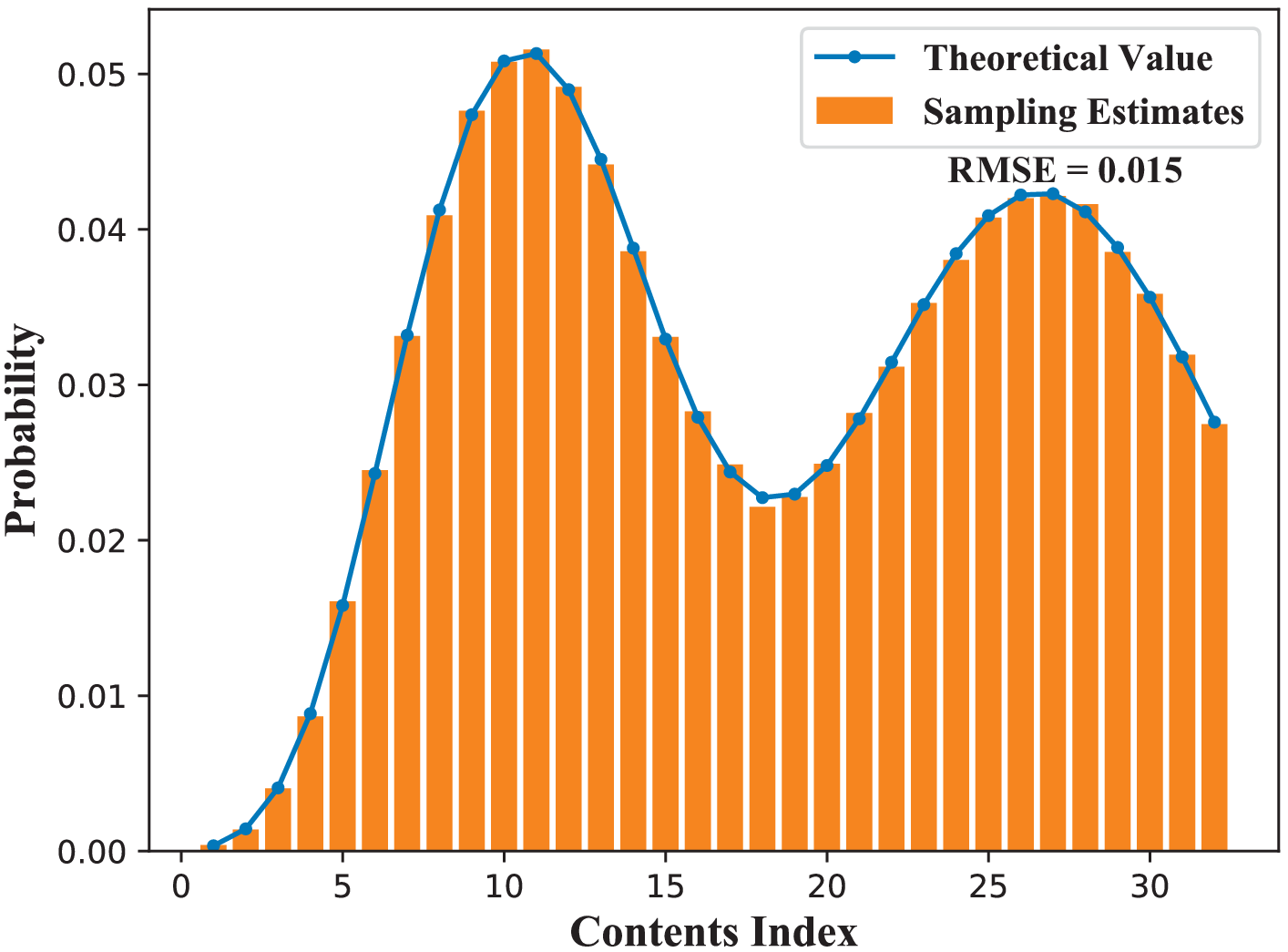}}%
    \hfil
    \subfigure[]{\includegraphics[width=0.5\columnwidth]{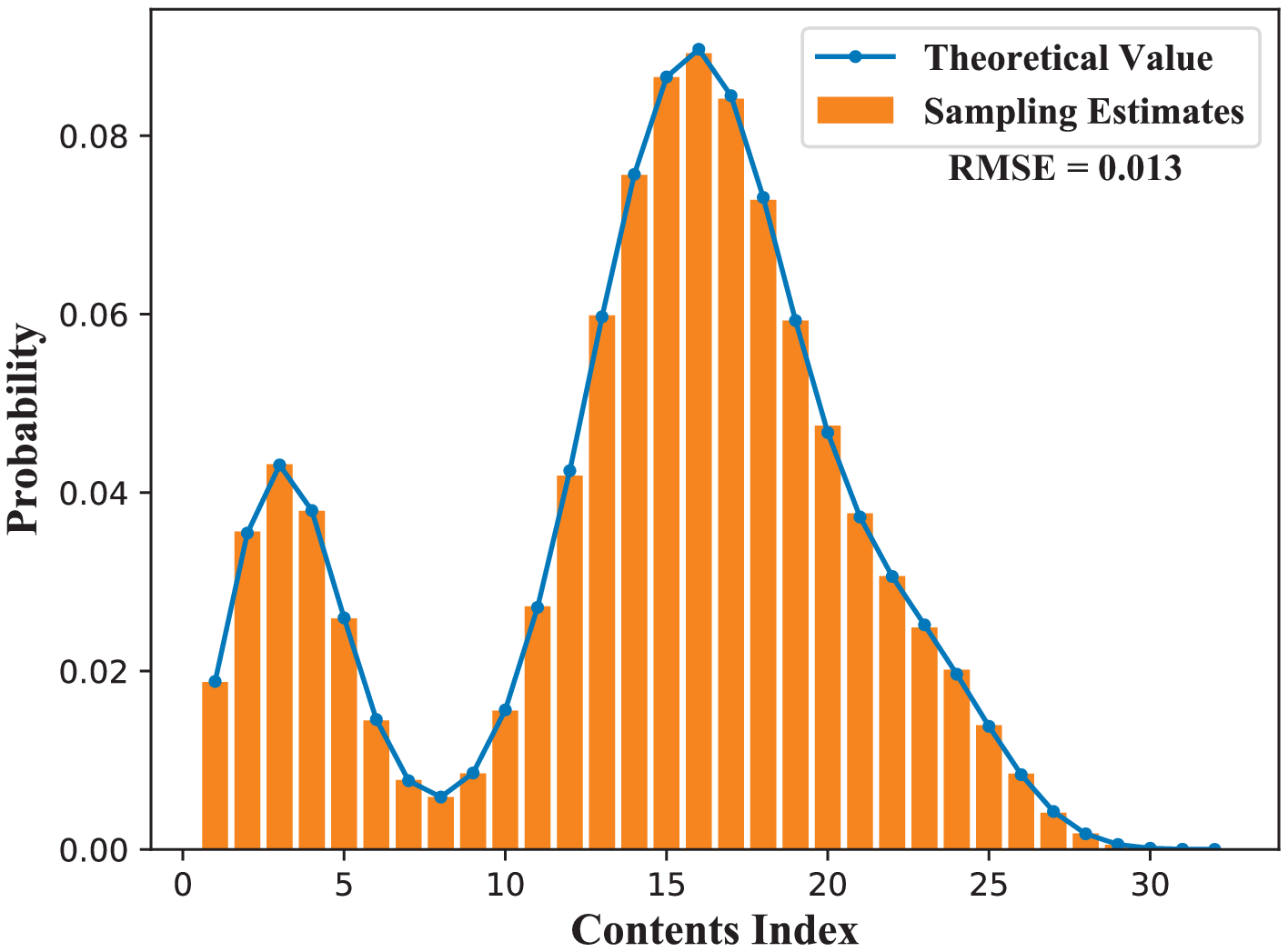}}%
    \hfil
    \subfigure[]{\includegraphics[width=0.5\columnwidth]{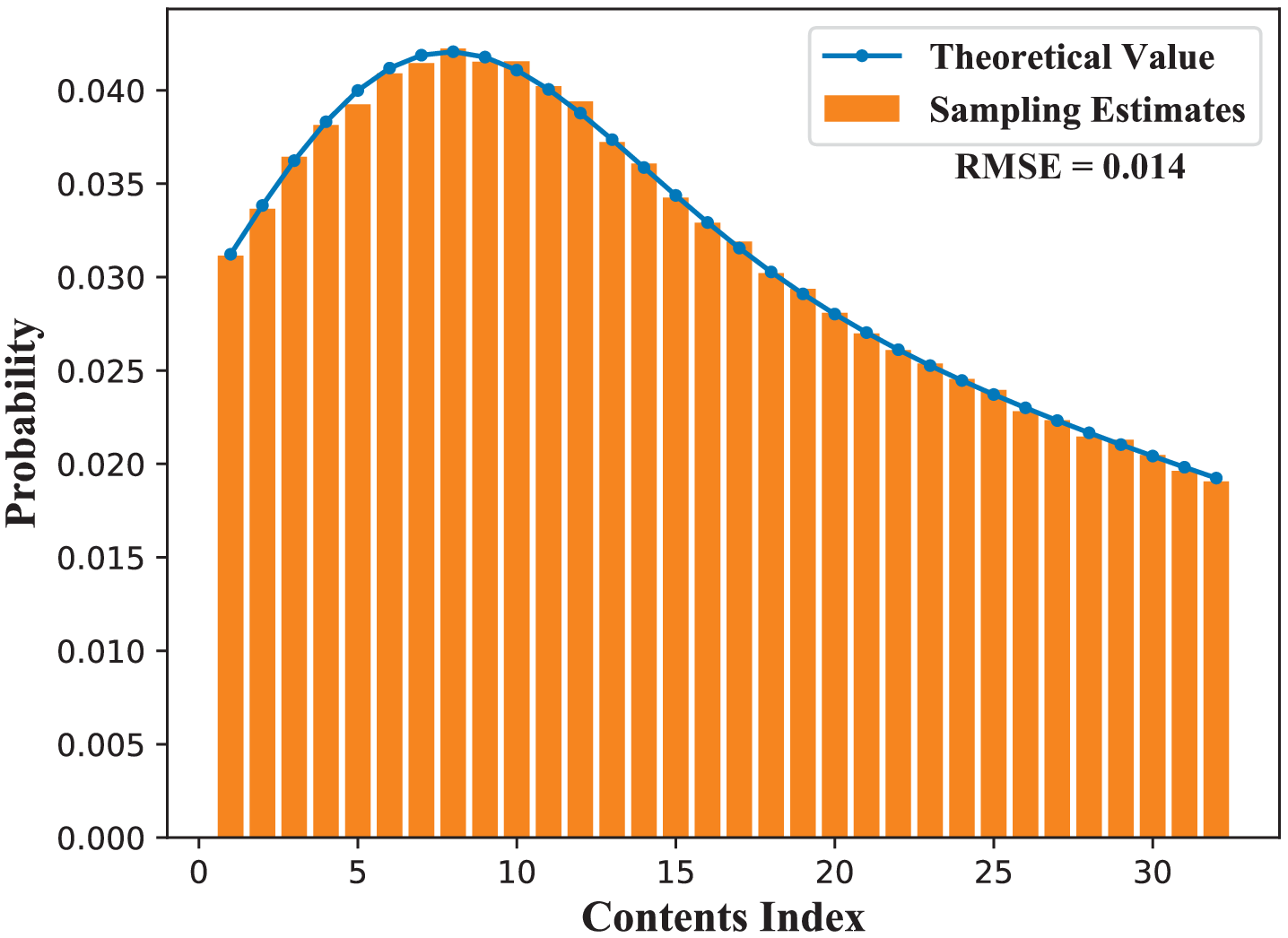}}
    \caption{Simulation validation of Theorem \ref{theorem1} on different probability distributions. ($I = 6$, $N=32$) (a) The local popularity of user $i$ follows ${\rm{Zipf}}\left(\alpha^{i}\right)$. (b) The local popularity of user $i$ follows $\rm{Poisson}\left(l^{i}\right)$. (c) The local popularity of user $i$ follows $\rm{nBernoulli}\left(p^{i}\right)$. (d) The local popularity of user $i$ follows $\rm{Gaussian}\left(\mu^{i},\sigma^{i}\right)$.}
    \label{fig_Theor1}
\end{figure*}

\section{CONCLUSION AND FUTURE WORK}
\label{sec5}

In this article, we investigated the problem of edge popularity prediction in a MEC-enabled privacy-sensitive IIoT system. The concepts of local popularity and global popularities are introduced and we reformulate the underlying distributed history-inaccessible time series forecasting problem as a label-absent distributed learning problem. The dynamic temporal dependencies within the long sequence are explored using LSTM cells and the training data labeling is circumvented by incorporating the AE structure. To realize collaborative prediction, the FL framework is adopted to effectively exchange the diverse model parameters of each network participant with a data-security guarantee. The above modules and designs collectively constitute a novel URFL algorithm, which achieves superior performance in terms of RMSE prediction error and AE loss while avoids privacy disclosure. Our future work will concentrate on the popularity prediction-assisted proactive caching, as well as more complicated scenarios such as the heterogeneous multiple MEC nodes and non-i.i.d. user behaviors.

\appendices
\section{PROOF OF THEOREM \ref{theorem1}}\label{theorem1proof}

Suppose that the occurrence of events ${F^i}(t) = {F_n}$ and ${F^i}(t) \notin \emptyset$ are respectively denoted by ${B^i}$ and ${C^i}$. According to the service process described in Section~\ref{ser_pro}, we readily have $\overline {{C^j}}  \cap {B^j} = \emptyset $, $B^{i}\subseteq C^{i}$, and the statuses of different users are mutually independent. From the perspective of the MEC server, all the local users can be treated as a whole. As such, if we let ${F^{\rm{G}}}(t)$ denote the possible request received at time slot $t$, we obtain the following derivation:
\begin{equation} \label{theorem1proof1}
\begin{array}{l}
P_{n}^{\textrm{G}}\left(t\right)=P\left\{ F^{\textrm{G}}\left(t\right)=F_{n}\right\} =\frac{P\left\{ \left.F^{\textrm{G}}\left(t\right)=F_{n}\right|F^{\textrm{G}}\left(t\right)\notin\emptyset\right\} }{P\left\{ F^{\textrm{G}}\left(t\right)\notin\emptyset\right\} }\\
=\frac{{P\left\{ {\left\{ {{F^{\rm{G}}}\left( t \right) = {F_n}} \right\} \cap \left\{ {{F^{\rm{G}}}\left( t \right) \notin \emptyset } \right\}} \right\}}}{{{{\left[ {P\left\{ {{F^{\rm{G}}}\left( t \right) \notin \emptyset } \right\}} \right]}^2}}} \vspace{1ex}\\
=\frac{{P\left\{ {\left\{ {\overline {\bigcap\nolimits_{j = 1}^I {\overline {{B^j}} } } } \right\} \cap \left\{ {\overline {\bigcap\nolimits_{i = 1}^I {\overline {{C^i}} } } } \right\}} \right\}}}{{{{\left[ {P\left\{ {\overline {\bigcap\nolimits_{i = 1}^I {\overline {{C^i}} } } } \right\}} \right]}^2}}} = \frac{{P\left\{ {\left\{ {\bigcup\nolimits_{j = 1}^I {{B^j}} } \right\} \cap \left\{ {\bigcup\nolimits_{i = 1}^I {{C^i}} } \right\}} \right\}}}{{{{\left[ {P\left\{ {\bigcup\nolimits_{i = 1}^I {{C^i}} } \right\}} \right]}^2}}} \\
= \frac{{P\left\{ {\bigcup\nolimits_{j = 1}^I {\bigcup\nolimits_{i = 1}^I {\left( {{B^j} \cap {C^i}} \right)} } } \right\}}}{{{{\left[ {P\left\{ {\bigcup\nolimits_{i = 1}^I {{C^i}} } \right\}} \right]}^2}}} = \frac{{\sum\nolimits_{j = 1}^I {\sum\nolimits_{i = 1}^I {P\left\{ {{B^j} \cap {C^i}} \right\}} } }}{{{{\left[ {\sum\nolimits_{i = 1}^I {P\left\{ {{C^i}} \right\}} } \right]}^2}}}\\
={{\sum\limits_{j = 1}^I {\sum\limits_{i = 1}^I {\left[ {{\lambda _i}\left( t \right) \cdot {\lambda _j}\left( t \right)P_n^j\left( {{\alpha ^j}\left( t \right),t} \right)} \right]} } } \mathord{\left/
 {\vphantom {{\sum\limits_{j = 1}^I {\sum\limits_{i = 1}^I {\left[ {{\lambda _i}\left( t \right) \cdot {\lambda _j}\left( t \right)P_n^j\left( {{\alpha ^j}\left( t \right),t} \right)} \right]} } } {{{\left[ {\sum\limits_{i = 1}^I {{\lambda _i}\left( t \right)} } \right]}^2}}}} \right.
 \kern-\nulldelimiterspace} {{{\left[ {\sum\limits_{i = 1}^I {{\lambda _i}\left( t \right)} } \right]}^2}}}\\
={{\sum\nolimits_{j = 1}^I {\left[ {{\lambda _j}\left( t \right)P_n^j\left( {{\alpha ^j}\left( t \right),t} \right)} \right]} } \mathord{\left/
 {\vphantom {{\sum\nolimits_{j = 1}^I {\left[ {{\lambda _j}\left( t \right)P_n^j\left( {{\alpha ^j}\left( t \right),t} \right)} \right]} } {\sum\nolimits_{i = 1}^I {{\lambda _i}\left( t \right)} }}} \right.
 \kern-\nulldelimiterspace} {\sum\nolimits_{i = 1}^I {{\lambda _i}\left( t \right)} }}.
\end{array}
\end{equation}

In the light of the above result, the global popularity can be computed as
\begin{equation} \label{theorem1proof2}
\begin{array}{l}
{{\bf{P}}^{\rm{G}}}(t) = {{\left\{ {\sum\nolimits_{i = 1}^I {\lambda_{i}\left(t\right) \cdot P_n^i({\alpha ^i}(t),t)} } \right\}_{n = 1}^N} \mathord{\left/
 {\vphantom {{\left\{ {\sum\nolimits_{i = 1}^I {\lambda_{i}\left(t\right) \cdot P_n^i({\alpha ^i}(t),t)} } \right\}_{n = 1}^N} {\sum\nolimits_{i = 1}^I {\lambda_{i}\left(t\right)} }}} \right.
 \kern-\nulldelimiterspace} {\sum\limits_{i = 1}^I {\lambda_{i}\left(t\right)} }}\\
= {{\sum\nolimits_{i = 1}^I {\lambda_{i}\left(t\right) \cdot \left\{ {P_n^i({\alpha ^i}(t),t)} \right\}_{n = 1}^N} } \mathord{\left/
 {\vphantom {{\sum\nolimits_{i = 1}^I {\lambda_{i}\left(t\right) \cdot \left[ {P_n^i({\alpha ^i}(t),t)} \right]_{n = 1}^N} } {\sum\nolimits_{i = 1}^I {\lambda_{i}\left(t\right)} }}} \right.
 \kern-\nulldelimiterspace} {\sum\nolimits_{i = 1}^I {\lambda_{i}\left(t\right)} }}\\
= {{\sum\nolimits_{i = 1}^I {\lambda_{i}\left(t\right) \cdot {{\bf{P}}^i}\left( {{\alpha ^i}(t),t} \right)} } \mathord{\left/
 {\vphantom {{\sum\nolimits_{i = 1}^I {\lambda_{i}\left(t\right) \cdot {{\bf{P}}^i}\left( {{\alpha ^i}(t),t} \right)} } {\sum\nolimits_{i = 1}^I {\lambda_{i}\left(t\right)} }}} \right.
 \kern-\nulldelimiterspace} {\sum\nolimits_{i = 1}^I {\lambda_{i}\left(t\right)} }}.
\end{array}
\end{equation}
\begin{flushright}$\blacksquare$\end{flushright}

\section{Simulation Validation OF THEOREM \ref{theorem1}}\label{theo1_simu}
Herein, we provide a statistical experiment to compare the gap between the sampling estimate and the theoretical value so as to further validate the Theorem \ref{theorem1}, where the sampling estimate of ${{\bf{P}}^{\rm{G}}}(t)$ is counted from the actual samples in the system and the theoretical value of ${{\bf{P}}^{\rm{G}}}(t)$ is computed using Theorem \ref{theorem1}.

As for the sampling estimate, we set up 6 users and respectively record their requests ${F_i}(t)$ over continuous 100000 time slots, where the parameter set $\left\{ \alpha^{i}(t),\lambda_{i}\left(t\right)\right\} $ of each user are randomly generated and remain constant during this period. Besides, the number of total contents $N$ is set to 32. Then, we count the number of times that each content $F_n\in\left\{ F_1,F_2,\cdots,F_N\right\}$ has been requested according to the recorded requests data. Finally, according to Borel's law of large numbers \cite{Sade17},  we observe the requested frequency of each content and acquire the sampling estimate of ${{\bf{P}}^{\rm{G}}}(t)$ by approximating ${{\bf{P}}^{\rm{G}}}(t)$ from these frequencies. On the other hand, the theoretical value of ${{\bf{P}}^{\rm{G}}}(t)$ can be directly calculated by Theorem 1 under this given scenario. Moreover, in the experiment, we consider the case that the local popularity follows four different probability distributions respectively, i.e., ${\rm{Zipf}}\left(\alpha^{i}\right)$, $\rm{Poisson}\left(l^{i}\right)$, $\rm{nBernoulli}\left(p^{i}\right)$, $\rm{Gaussian}\left(\mu^{i},\sigma^{i}\right)$. $\alpha^{i}$, $l^{i}$, $p^{i}$, $\{\mu^{i},\sigma^{i}\}$ respectively represent the distribution parameter of user $i$ under these four probability distributions. Specifically, the parameter settings in this statistical experiment are listed in Table \ref{statis_set}. From Fig. \ref{fig_Theor1}, we can observe that the gap between the sampling estimate and the theoretical value always stays negligible under different distributions, which validates the Theorem \ref{theorem1} intuitively. Besides, as shown in Fig. \ref{fig_Theor1}, the gap is also quantified by the RMSE metric to further validates the Theorem \ref{theorem1}.

%\ifCLASSOPTIONcaptionsoff
%  \newpage
%\fi
%
\balance

\end{document}